\documentclass[preprint]{aastex631}

\newcommand{\mafont}{
  \bfseries
  \color{blue}
}

\usepackage{hyperref}
\usepackage{soul}
\DeclareTextFontCommand{\ma}{\mafont}
\usepackage[normalem]{ulem}
\newcommand\maout{\bgroup\markoverwith{\textcolor{blue}{\rule[.5ex]{2pt}{1.0pt}}}\ULon}

\begin{document}
%\tableofcontents

\title{The Role of Magnetic Shear in Reconnection-Driven Flare Energy Release}

\author{J. Qiu}
\affiliation{Department of Physics,
Montana State University,
Bozeman, MT 59715, USA}

\author{M. Alaoui}
\affiliation{IREAP,
University of Maryland,
College Park, MD 20742, USA}

\author{S. K. Antiochos}
\affiliation{IREAP,
University of Maryland,
College Park, MD 20742, USA}

\author{J. T. Dahlin}
\affiliation{
Astronomy Department,
University of Maryland, College Park, MD 20742, USA}

\author{M. Swisdak}
\affiliation{IREAP,
University of Maryland,
College Park, MD 20742, USA}

\author{J. F. Drake}
\affiliation{IREAP,
University of Maryland,
College Park, MD 20742, USA}

\author{A. Robison}
\affiliation{Department of Physics,
Pepperdine University,
24255 Pacific Coast Highway,
Malibu, CA 90263, USA}

\author{C. R. DeVore}
\affiliation{Heliophysics Science Division, 
NASA Goddard Space Flight Center,
8800 Greenbelt Road,
Greenbelt, MD 20771, USA}

\author{V. M. Uritsky}
\affiliation{Catholic University of America at NASA Goddard Space Flight Center,
8800 Greenbelt Road,
Greenbelt, MD 20771, USA}


\begin{abstract}

Using observations from the Solar Dynamics Observatory's Atmosphere Imaging Assembly and the Ramaty High Energy Solar Spectroscopic Imager, we present novel measurements of the shear of post-reconnection flare loops (PRFLs) in SOL20141218T21:40 and study its evolution with respect to magnetic reconnection and flare emission. Two quasi-parallel ribbons form adjacent to the magnetic polarity inversion line (PIL), spreading in time first parallel to the PIL and then mostly in a perpendicular direction. 
%We measure the shear angle %($\theta$) of a large number of PRFLs as the complement of the angle of these loops observed in extreme ultraviolet passbands ($\lesssim$1 MK) with respect to the PIL. 
We measure magnetic reconnection rate from the ribbon evolution, and also the shear angle of a large number of PRFLs observed in extreme ultraviolet passbands ($\lesssim$1 MK).
For the first time, the shear angle measurements are conducted using several complementary techniques allowing for a cross-validation of the results. In this flare, the total reconnection rate is much enhanced before 
a sharp increase of the hard X-ray emission, and the median shear decreases from 60$^\circ$-70$^\circ$ to 20$^\circ$, on a time scale of ten minutes. 
We find a correlation between the shear-modulated total reconnection rate and the non-thermal electron flux. These results confirm the strong-to-weak shear evolution suggested in previous observational studies and reproduced in numerical models,  
and also confirm that, in this flare, reconnection is not an efficient producer of energetic non-thermal electrons during the first ten minutes when the strongly sheared PRFLs are formed. %, nor 
%{\vu No energetic electron are seen} 
%in the late flare phase when the shear of the PFRLs nears zero. 
We conclude that an intermediate shear angle, $\le 40^\circ$, is needed for efficient particle acceleration via reconnection, and we propose a theoretical interpretation.

\end{abstract}


\section{Introduction} \label{sec:intro}

Magnetic reconnection in the solar corona is widely believed to be the energy release mechanism that drives solar flares.  For eruptive two-ribbon flares, the Carmichael–Sturrock–Hirayama–Kopp–Pneuman (CSHKP) model \citep{Carmichael1964, Sturrock1966, Hirayama1974, Kopp1976} provides the canonical description.  An arcade of flare loops form: at their foot-points, two flare ribbons spread apart and away from the magnetic polarity inversion line (PIL) as reconnection proceeds along a vertical current sheet in the corona. The 
%direction of the current follows the PIL and the% it does? why?
leading edges of the ribbons map the feet of the reconnecting magnetic field lines \citep{Svestka1980, Forbes1984}. The model also schematically describes the evolution of energized particles and plasma, as well as the dynamics of the lower atmosphere in response to the flare energy deposition.

The greatest challenge to understanding flare reconnection is that it occurs in the corona, where detailed, accurate measurements of the magnetic field are very rare. The standard flare model connecting the dynamics in the corona to the lower atmosphere response, however, provides a recipe for inferring reconnection properties by tracking the evolution of flare ribbons. For a typical Alfv\'en speed of order 1,000 km s$^{-1}$ and length scale of 10,000 km, the reconnection-released energy flux travels along the flare loops, i.e.,\ the closed field lines formed by reconnection, to reach and heat the upper chromosphere in a matter of seconds. Therefore, signatures of impulsive brightening in the lower atmosphere may be tracked to derive the reconnected flux, $\psi = \int B_r da $, where $B_r$ is the photospheric radial magnetic flux density and $da$ is the area of newly brightened flare ribbons. Its time derivative $\dot{\psi}$ gives the global reconnection rate. For strictly two-dimensional models such as CSHKP, the global reconnection rate is equivalent to a uniform reconnection electric field $E_{rec} = \dot{\psi}/L = v_{r}B_{r}$, where $L$ is the length of the macroscopic reconnection current sheet (RCS) running along the axis of the arcade and $v_{r}$ is the apparent speed of the ribbons perpendicular to the PIL \citep{Forbes1984, Forbes2000}. 
%T The magnetic flux encompassed by the growing ribbons acts as a proxy for the flux participating in reconnection in the corona 
%The global reconnection rate, defined here as $\dot{\psi}$, describes the total amount of flux participating in reconnection per unit time. 
The reconnection rate, in terms of $\dot{\psi}$ or $E_{rec}$, has been measured in this way for more than two decades \citep{Poletto1986, Fletcher2001, Qiu2002, Qiu2004, Isobe2002, Isobe2005, Krucker2003, Jing2005, Saba2006, Temmer2007, Liu2009b, Kazachenko2017, Hinterreiter2018}.
 
In essentially all models of flares, including CSHKP, the ultimate energy source for the event is the magnetic free energy stored in the strongly sheared field of a filament channel \citep[e.g.][]{Patsourakos20}. The basic scenario for eruptive two-ribbon flares is that the eruption ejects the shear, after which reconnection relaxes the field toward a potential state. Consequently, the flare reconnection is presumed to start between field lines that are not anti-parallel. An invariant component of the inflow magnetic field, often called the guide field or shear component $B_g$, is expected to vary as the location of the reconnecting field, $B_{rec}$, rises in altitude. Post-reconnection flare loops (PRFLs) are also expected to make an angle with the PIL, an angle that varies during the flare, as has been demonstrated in observations \citep[e.g.][]{Aschwanden2001}. The shear variation is also manifest in the apparent motions of flare ribbons or kernels, observed in the optical, ultraviolet, and hard X-ray (HXR) emissions that map the feet of the PRFLs. Observations have often shown that flare ribbons or kernels at first move or spread along the PIL and then move away from it \citep{Vorpahl1976, Kawaguchi1982, Kitahara1990, Krucker2003, Fletcher2004, Bogachev2005, Lee2008, Yang2009, Qiu2009, Qiu2010, Qiu2017}. For the along-the-PIL motion, conjugate flare foot-points in magnetic fields of opposite polarities may move in the same direction (i.e., zipper motion) or in opposite directions, either approaching or receding from each other.  The parallel-to-perpendicular evolution of this motion is sometimes related to changes in shear, the angle made by the line connecting conjugate foot-points with respect to 
%the vertical of (?)
the PIL. For two decades, observations have revealed strong-to-weak shear evolution in two-ribbon flares \citep{Aschwanden2001, Su2006, Ji2006, Su2007, Liu2009a, Yang2009, Qiu2010, Qiu2017, Qiu2022}, suggesting that the relative guide field, defined by $\mathcal{R} \equiv B_g/B_{rec}$, in the RCS decreases during flare reconnection. Note that strong-to-weak shear evolution is not necessarily present in all flares. Many flares exhibit irregular motions of the conjugate foot-points \citep[e.g.][]{Fletcher2004, Bogachev2005, Grigis2005, Yang2009, Cheng2012, Inglis2013}, reflecting the complex configurations or tempo-spatial structures of flare reconnection.

Reconnection releases magnetic energy, a significant amount of which is transferred to non-thermal particles \citep{Emslie2012, Aschwanden2019}. Past observational studies have often shown that HXR (or microwave) emissions are temporally, and sometimes spatially, correlated with $\dot{\psi}$ or $E_{rec}$ \citep{Qiu2002, Qiu2004, Krucker2003, Fletcher2004, Lee2006, Temmer2007, Jing2007}, albeit sometimes with delays in the HXRs on the order of 1-2 minutes \citep[e.g.][]{Miklenic2007, Yang2011, Naus2022, Vievering2023}. 
On the other hand, most of these studies did not verify a one-to-one coincidence between significant HXR emission and an enhanced reconnection rate, however the latter is measured. In particular, flare emissions in UV, optical, or soft X-rays (SXRs) and the inferred reconnection rates, may rise well before the occurrence of impulsive and significant non-thermal HXRs \citep[see][for some prominent examples]{Warren2001, Su2006, Krucker2011, Caspi2014, Naus2022}, and it has not been clear what mechanisms govern the partition of flare energy during different stages of the flare evolution. 

Recent numerical simulations find that flare energetics depend critically on the reconnection rate as well as the guide field \citep{Dahlin2015, Arnold2021}. The models predict that the ratio of the guide field to the reconnecting component plays an important role in determining the efficiency of particle acceleration via magnetic islands \citep{Dahlin2017,Dahlin2020,Arnold2021}. Consequently, experimental determination of the guide field could provide stringent tests of the theoretical models. 
Furthermore, while the phenomenological relationship between the reconnection rate, $\dot{\psi}$ or $E_{rec}$, and flare emission has been intensively studied in observations, the role of $B_g$ has not been considered because of the difficulty of quantifying this parameter. In this paper, we 
%attempt to 
measure the shear of post-reconnection flare loops (PRFLs) as a proxy for the relative guide field $\mathcal{R} = B_g/B_{rec}$ during the evolution of the M6.9 two-ribbon flare SOL20141218T21:40. We then investigate how the shear is related to the flare energetics, in particular the efficiency of converting free magnetic energy into kinetic energy of non-thermal electrons. 
We also relate the evolution of shear in the observed flare to the evolution of shear and reconnection guide field in a three-dimensional simulation of an eruptive solar flare \citep{Dahlin2022}.

Our paper is organized as follows. Section \ref{sec:overview} provides an overview of the flare observations. Section \ref{sec:motion} discusses the evolution of the flare ribbons and loop tops, and Section \ref{sec:shear} the shear evolution of the post-reconnection flare loops.  The flare energetics are the focus of Section \ref{sec:HXR}. 
Inferences from recent modeling work that yield insight into our results are developed in Section 6.
Section \ref{sec:summary} offers a summary of our findings and final conclusions.

\section{Overview of the SOL20141218T21:40 M6.9 Two-ribbon Flare} \label{sec:overview}

\begin{figure}    %%%%%%%%%%%%%%%%%% 
\centerline{\includegraphics[width=1.0\textwidth,clip=]{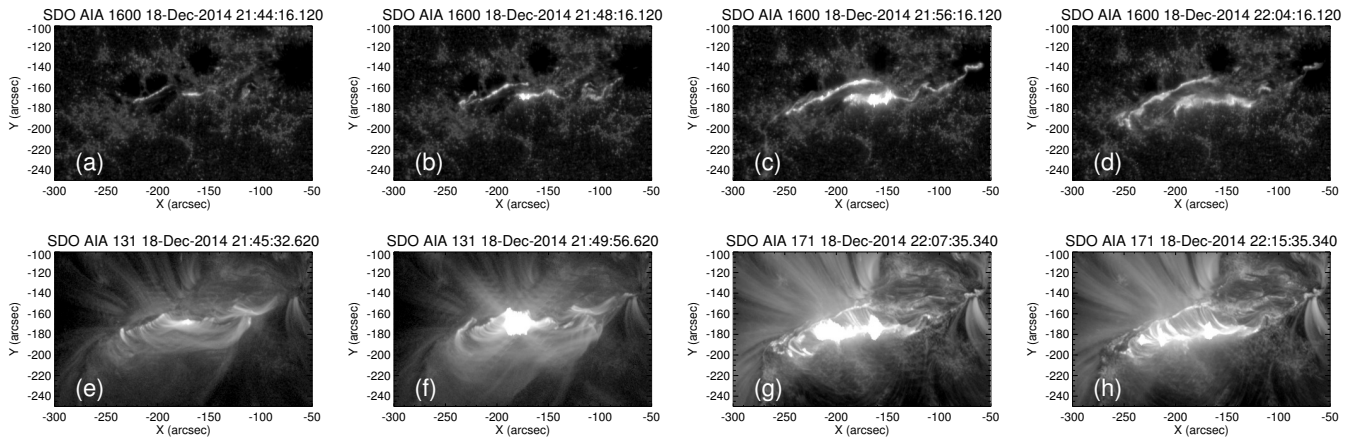}
              }         
	\caption{Overview of the SOL20141218T21:40 M6.9 flare. (a-d) Evolution of flare ribbons observed 
 in UV 1600~\AA\ by AIA. (e-h) Post-reconnection flare loops (PRFLs) observed in the EUV 131 or 171~\AA\ passbands by AIA. Images in all panels have been rotated to 21:00 UT and therefore co-aligned. }%{\jqfont to be expanded with a blow-up magnetogram, (and w/ measurements of mean field in P and N ribbons, like shown before?)}
 \label{fig:overview}  
   \end{figure}

The M6.9 flare discussed here occurred in the active region NOAA AR 12241 and was accompanied by a coronal mass ejection (CME). \citet{Joshi2017} studied its evolution in great detail. They suggested that the CME flux rope was formed in-situ in the early phase of the flare by reconnection of a sheared arcade.   
% \citep[also see][]{Prasad2022}. %prior to 21:50~UT
 The rope erupted as soon as it formed, accelerating the reconnection and forming an arcade of post-reconnection flare loops (PRFLs) anchored to two parallel flare ribbons along the PIL (Figure~\ref{fig:overview}a-h). This scenario resembles ``tether-cutting" reconnection \citep{Moore2001} in a modified standard model. Subsequently, the erupting rope interacted (reconnected) with high-lying flux, forming a remote circular ribbon \citep[not shown in this paper; see Figures 10 and 11 of][]{Joshi2017} before finally escaping the corona, indicating that AR 12241 was in a ``breakout'' reconnection configuration \citep{Antiochos1999}. 
%MS: I don't think this sentence is needed: In this study, we focus on the evolution of the reconnection and energetics next to the PIL by analyzing flare ribbons and PRFLs observed by AIA as well as X-ray emissions observed by RHESSI.

\begin{figure}    %%%%%%%%%%%%%%%%%% 
        \centerline{\includegraphics[width=1.0\textwidth,clip=]{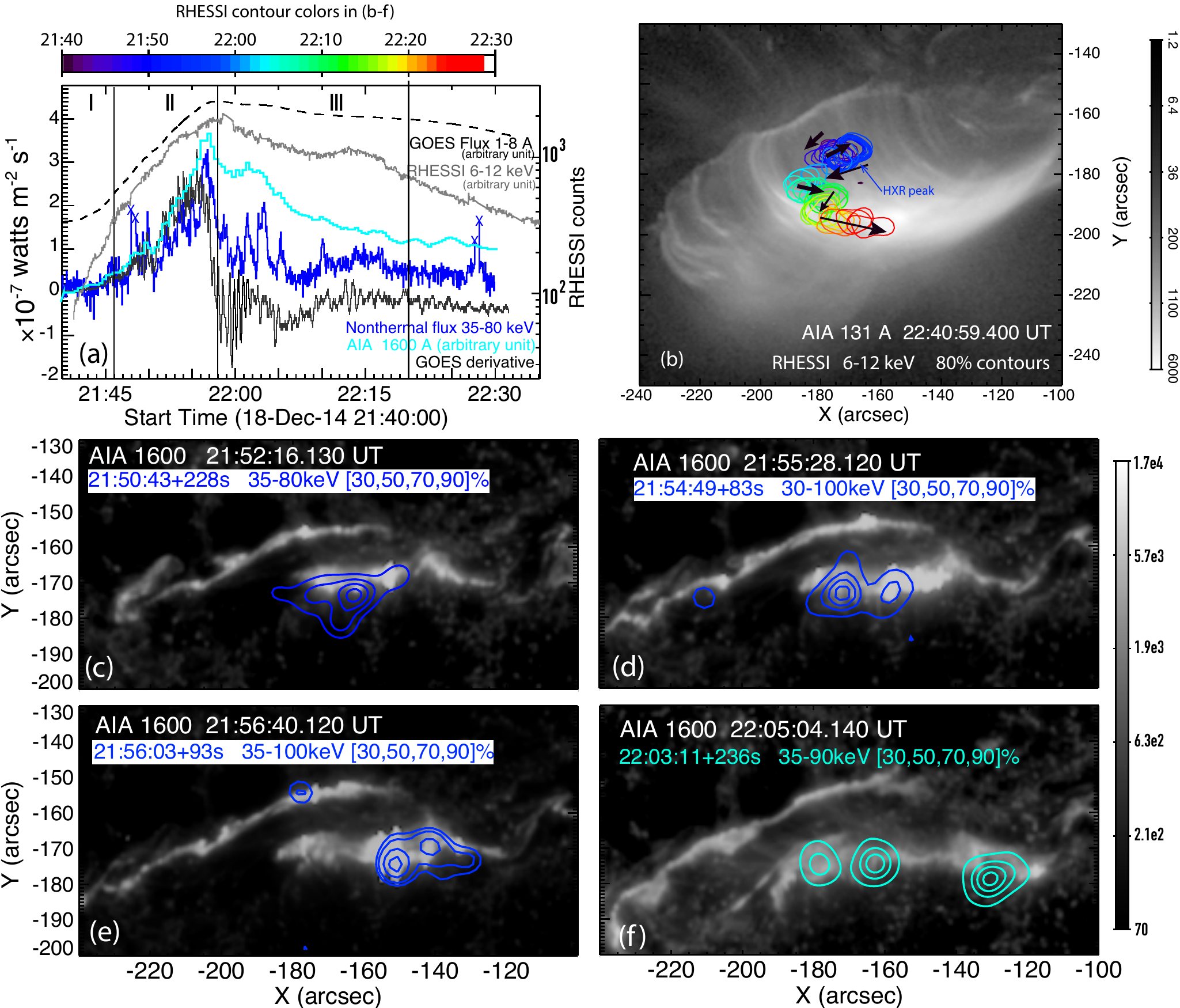}
              }
              \caption{(a) Soft X-ray (1-8~\AA) and its time derivative, hard X-ray (6-12 and 35-80~keV), and UV 1600~\AA\ light curves of the flare obtained by GOES, RHESSI, and AIA, respectively. The solid vertical lines divide the flare evolution into three stages (see text).
              %{\crdfont{[Move rightmost vertical line in (a) from time 22:05 to time 22:20. -- done]}}
              (b) RHESSI X-ray 6-12~keV 80\% contours plotted over an AIA 131~\AA\ image at 22:40:59 UT. %The colors represent the middle of the image integration time. 
              The colors indicate the midpoint of the RHESSI integration time interval; see color bar above panel (a). The black arrows indicate the apparent motion of the HXR centroids over time.
              (c-f) HXR $>30$ keV contours superimposed on AIA 1600~\AA\ images. RHESSI contour colors correspond to the color bar above panel (a) to easily follow the time evolution.} % {\jqfont Q for Meriem: 1. in (a), how the time derivative of GOES SXR is positive all the time? Shouldn't it become negative after the peak of GOES emission? 2. for (b), may we use a different AIA background image? The 193 image at 21:52 is a bit too early and do not show loops -- or do we see hot 20MK loops in this wavelength early on? An EUV image 10-20 minutes or more later would allow to see PRFLs and the 6-12 keV locations relative to these loops. }} 
               \label{fig:hxr}  
   \end{figure}

Figure~\ref{fig:overview} presents an overview of the evolution of the flare adjacent to the PIL. 
%but does not include its later development when the flux rope further interacts with overlying large-scale flux systems. 
The flare was observed by AIA with a time cadence of 24~s in the UV 1600 passband and 12~s in each of the seven EUV passbands.
The flare ribbon development follows the elongation-to-expansion style, with ribbons rapidly spreading along the PIL in the first few minutes, followed by expansion away from and perpendicular to the PIL, first rapidly, then more gradually. Based on the flare ribbon morphology (Figure~\ref{fig:overview}a-d), we track the flare evolution during the intervals marked in Figure~\ref{fig:hxr}a: 21:40-21:46 UT, 21:46-21:58 UT, and 21:58-22:20 UT, marked as phases I, II, and III, respectively. 
%The evolving two ribbons are also mapped in the photospheric radial magnetogram in Figure~\ref{fig:motion}a, with the color code indicating the time the ribbons start to brighten. 
% I think the above sentence breaks the rule of numbering figures in the order shown.
Such elongation-to-expansion development is often accompanied by the strong-to-weak shear evolution of PRFLs reported in many previous studies \citep{Aschwanden2001, Su2006, Su2007, Liu2009b, Qiu2010, Qiu2017, Qiu2022}, and is also evident in this flare as shown in the EUV images in Figure~\ref{fig:overview}e-h. 

Figure~\ref{fig:hxr}a illustrates the flare emission in  the soft X-ray (SXR) 1-8~\AA\ band observed by GOES, its time derivative, the X-rays at photon energies 6-12~keV and 35-80~keV observed by RHESSI, and the total count rates in the UV 1600~\AA\ passband from AIA. The three stages of the flare ribbon evolution are marked, showing that the first stage, the ribbon {\it elongation} stage, has little energetic consequence in terms of flare radiation. The second stage, the {\it fast expansion} stage, is coincident with the impulsive phase of the flare when emissions peak; during this stage non-thermal hard X-rays (HXRs), represented by $\ge 30$~keV emission, are most significant. Finally, the third stage, the {\it slow expansion} stage, is coincident with the decay phase of the flare, when flare emissions have passed their peak and $\ge$30~keV HXRs have decreased. Note that the flare SXR light curves do not exhibit a simple smooth decay after the peak, suggesting additional episodes of energy release, possibly involving reconnection between the erupting flux rope and overlying flux systems \citep{Joshi2017}. In these later episodes, however, energetic HXR emissions beyond 30~keV are significantly diminished.
%% This is not a universally accepted definition but SXR and HXR usually refer to the thermal and non-thermal parts of a spectrum, respectivelt.

X-ray images from RHESSI are constructed and displayed in Figure~\ref{fig:hxr}b-f. To generate these images, data from detectors 3 and 6 through 9 were reduced with the \textit{clean} algorithm using a beam width of 2 \citep{2009ApJ...698.2131D}, making the angular resolution $\approx$~7\arcsec. Images of the SXR sources between 6-12~keV were taken from the RHESSI archive.\footnote{\url{ https://hesperia.gsfc.nasa.gov/rhessi\_extras/flare\_images/hsi\_flare\_image\_archive.html}. The images are constructed by applying the CLEAN algorithm to data from detectors 3, 6, 7, 8 and 9, and the integration time of each map varies from 16~s to 120~s.} The SXR emission below 20~keV is located close to the southern ribbon, but not at the same place as the 30-80~keV sources; the lower-energy X-ray emission likely comes from or above the top of newly formed flare loops. The 6-12~keV source exhibits apparent motion initially along the PIL. %suggesting that the source is moving upward as the flare progresses. %\ma{There is no clear evidence of the LT source motion downward first, as reported in numerous events: Veronig et al 2006,}%%% XXX
After 22:05~UT, the source moves beyond the PRFL system observed in EUV passbands, possibly due to production at higher altitude when the erupting flux rope interacts with high-lying flux systems.

\begin{figure}    %%%%%%%%%%%%%%%%%% 
\includegraphics[width=0.55\textwidth,clip=]{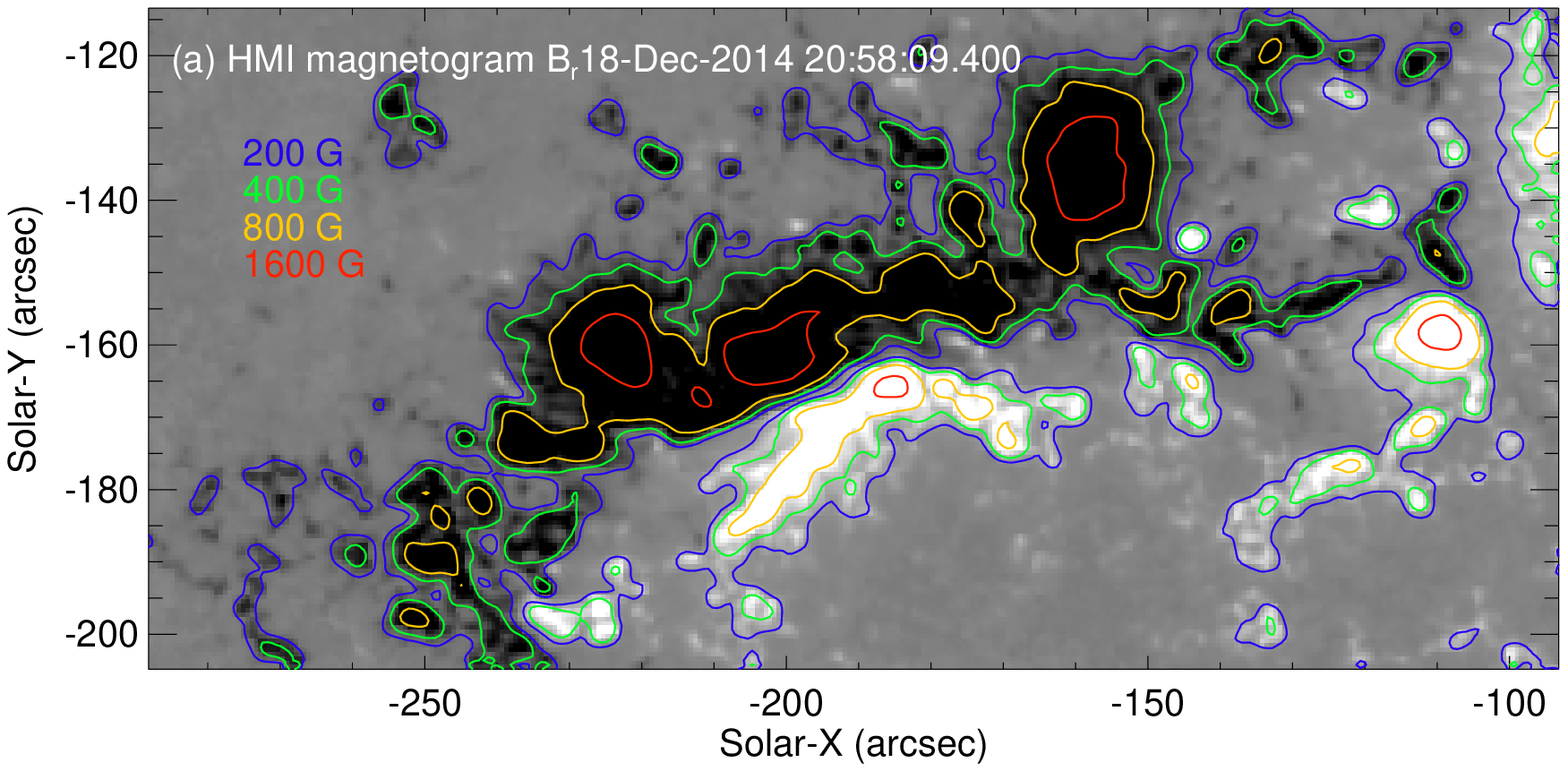}
\includegraphics[width=0.45\textwidth,clip=]{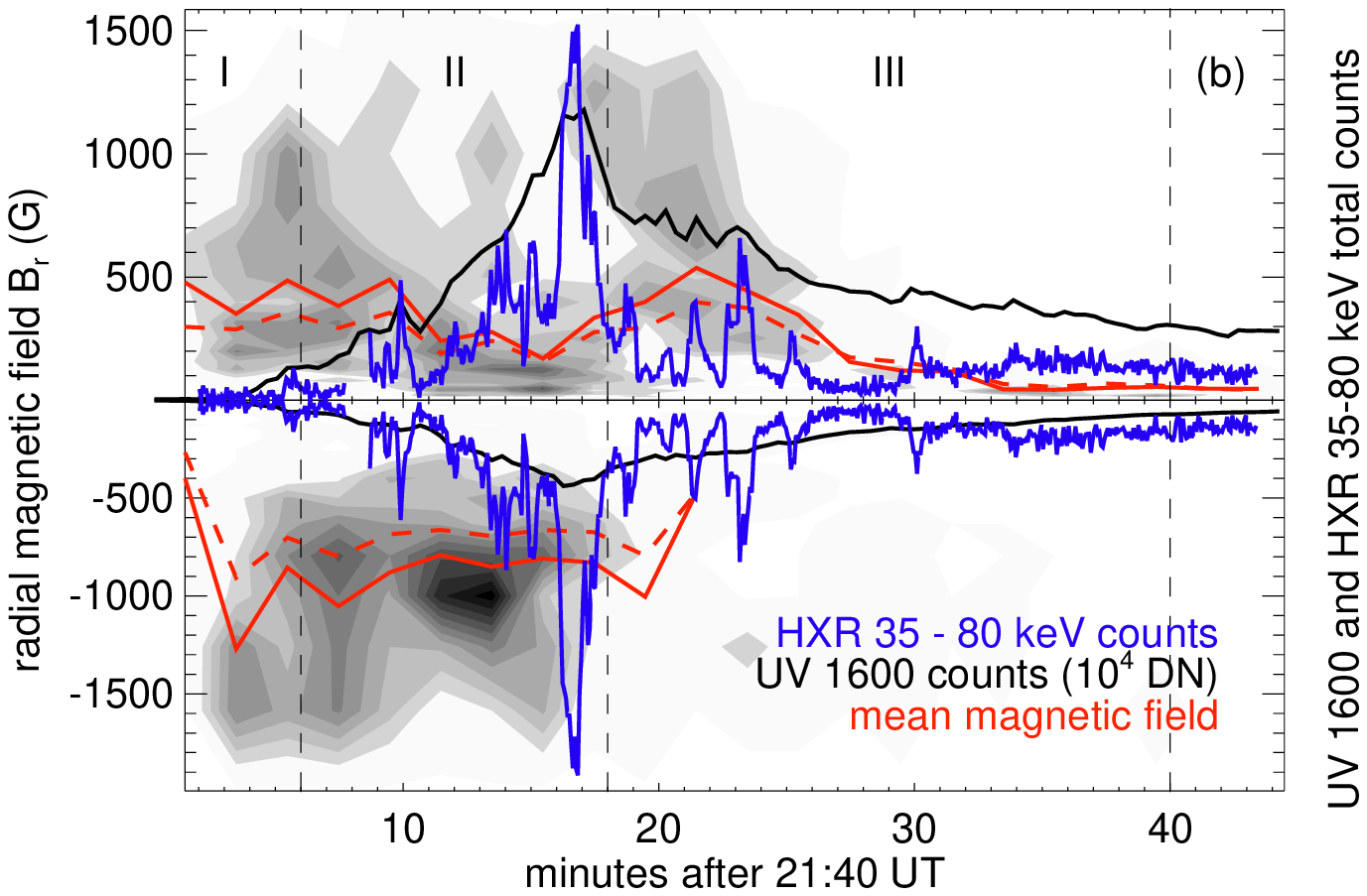}
	\caption{
 (a) Photospheric radial magnetogram obtained from HMI/SHARP. %{\vu PERHAPS WE COULD COLOR-CODE THE CONTOURS?} 
 Color contours indicate magnetic flux density of $\pm$200, 400, 800, 1600 Mx cm$^{-2}$. (b) 2D histogram of photospheric radial magnetic flux density (gray scale) on the southern (positive $B_r$) and northern (negative $B_r$) ribbons, respectively, integrated for every two minutes. Red curves give the mean magnetic field $\langle B_r \rangle$ on the photosphere (solid) and the mean magnetic field extrapolated to 1~Mm above the photosphere (dashed) on the ribbons. These are compared with the total HXR count rates at 35-80~keV (blue; arbitrary units), and the total UV emissions (black) integrated on the positive and negative ribbons separately.
 %{\crdfont{[Add dashed vertical line to panel (b) at time 40, showing the end of Phase III?  - done]}}
 }
\label{fig:mmap}  
   \end{figure}

The HXR sources in 30-100 keV are constructed between 21:50~UT and 22:05~UT, with integration times of 80-240~s to obtain good count statistics. These sources are mostly located at the southern ribbon and cover nearly all of it.
%\st{, consistent with the spectral fits and the thick-target model}. 
The presence of multiple HXR sources along the southern ribbon may be partially due to the dynamic range of the instrument. However, it is noteworthy that the locations of the sources coincide with co-temporal regions of increased flux in AIA 1600 \AA, indicating a higher energy deposition at discrete locations along the ribbon. We find very little thick-target HXR emission at the northern ribbon. Such an asymmetry in the HXR thick-target source has been often observed \citep{Sakao1994, Yang2009} and likely reflects a magnetic mirroring effect where the weaker HXR source corresponds to the region of higher photospheric magnetic field strength \citep{Melrose1981, Liu2009b, Daou2016}.  %{\jqfont (double check preceding references ..)}.% for statistical study that reported 2/3 of events show the stronger HXR foot-point is co-spatial with the weaker magnetic field ribbon). 
In support of the magnetic mirroring scenario, Figure~\ref{fig:mmap} shows the photospheric radial magnetogram in the flare active region (panel a), obtained by the Helioseismic and Magnetic Imager \citep[HMI;][]{Schou2012} and the Spaceweather HMI Active Region Patch (SHARP) database. The distribution of magnetic field strength (flux density) on each of the flare UV ribbons is displayed in gray scale in panel b. The mean magnetic flux density (red curves in panel b) on the northern ribbon in the negative magnetic field is more than twice that on the southern ribbon located in the positive field, and at the peak time the total UV emission on the northern ribbon is one third that on the southern ribbon.

This study focuses on the major phase of the flare development adjacent to the PIL, when and where flare emissions are most energetic. In the subsequent sections we derive properties of magnetic reconnection from the evolution of the flare ribbons and PRFLs and investigate how they are related to the flare energetics, in particular the non-thermal energetics reflected in the HXR emissions.

\begin{figure}    %%%%%%%%%%%%%%%%%% 
\includegraphics[width=0.49\textwidth,clip=]{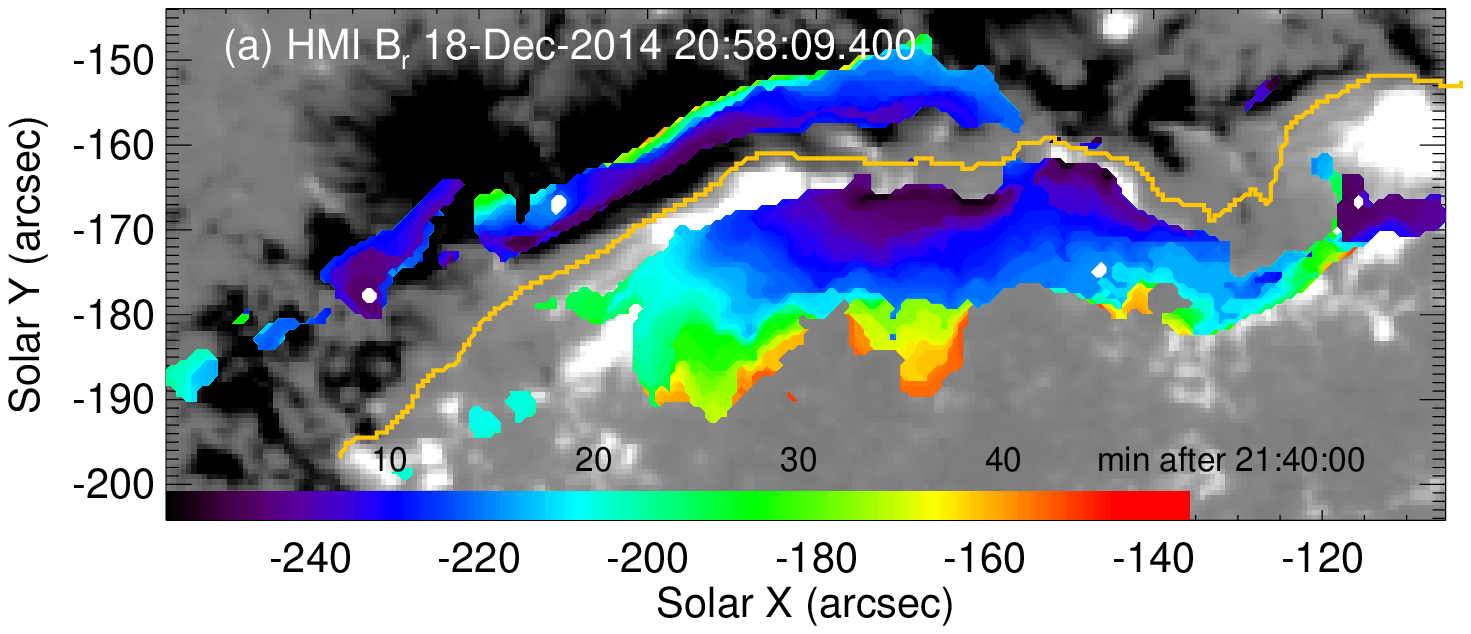}
\includegraphics[width=0.49\textwidth,clip=]{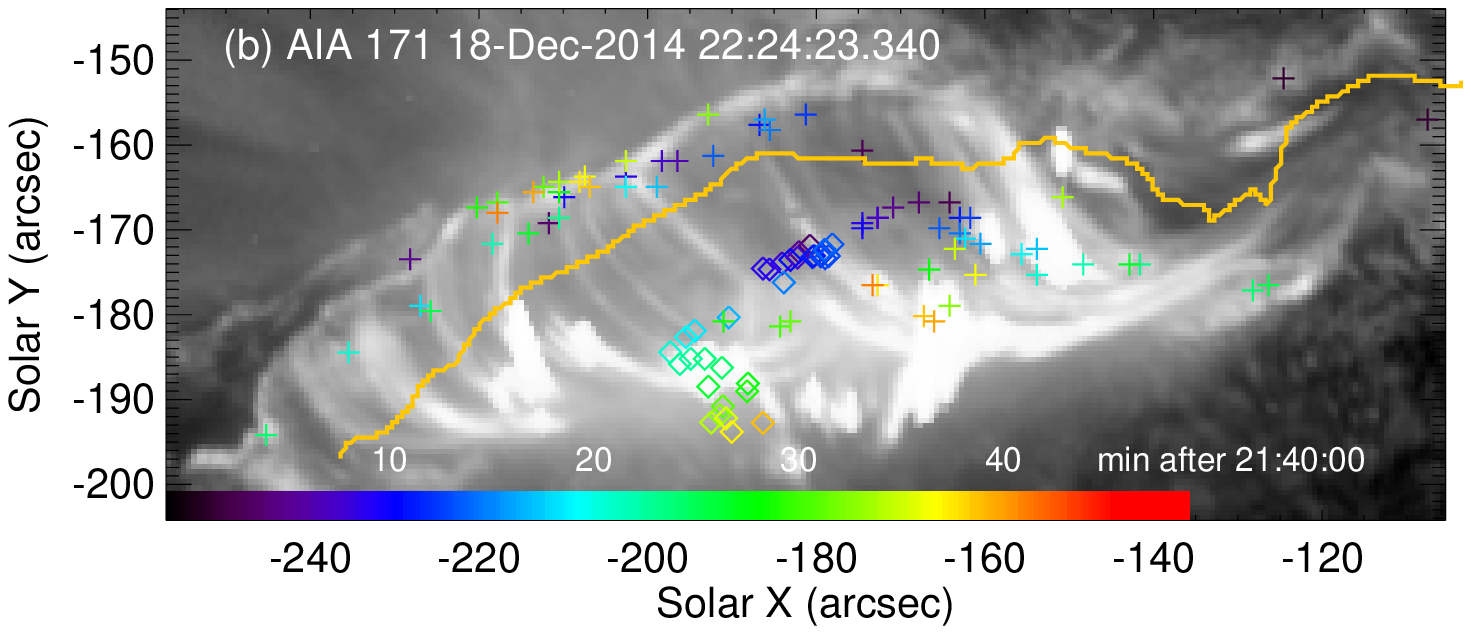}
\includegraphics[width=0.53\textwidth,clip=]{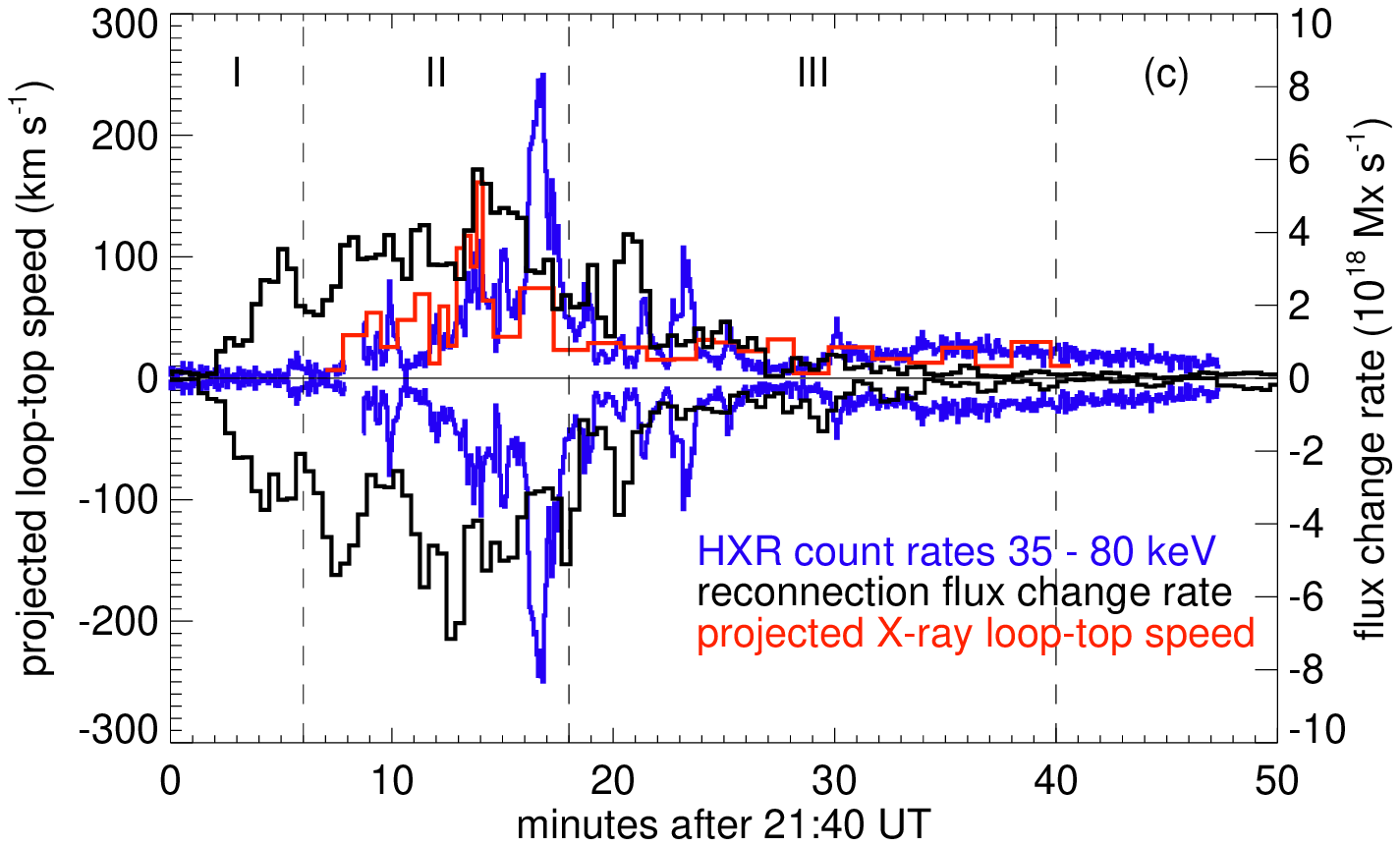}
\includegraphics[width=0.45\textwidth,clip=]{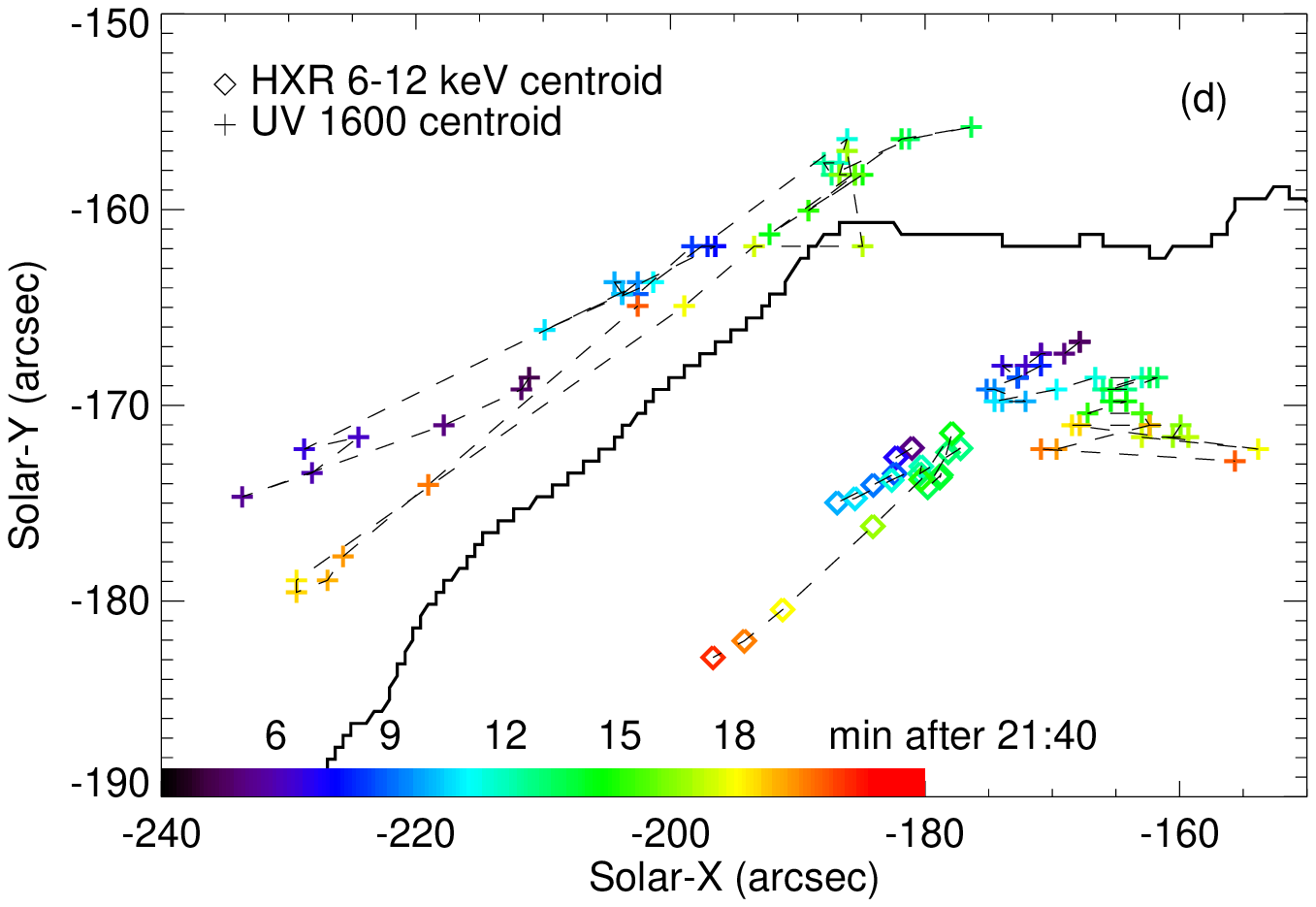}  
	\caption{
 (a) Evolution of the flare ribbon fronts, derived with the AIA 1600~\AA\ images, superimposed on a radial photospheric magnetogram obtained from HMI/SHARP.  
 (b) HXR 6-12~keV (RHESSI) centroids (diamonds; with $f = 0.85$, see text) and the UV 1600~\AA\ (AIA) centroids (pluses; with $f = 0.7$, see text) superimposed on an EUV image in the AIA 171~\AA\ passband.  For clarity of display, we have reduced the cadence of the UV centroids to 72~s, or every third frame.
 (c) Total reconnection rate in terms of the flux change rate $\dot{\psi}$ (black) measured in the positive field and negative field, respectively, the mean
plane-of-sky motion velocity of the HXR 6-12 keV centroids, computed with the entroids measured with $f = 0.75, 0.85, 0.95$ (red), and the HXR count rates at 35-80~keV (blue). %Red vertical bars indicate the range of $v_{top}$ computed with the centroids measured with $f = 0.75, 0.85, 0.95$ respectively. 
 (d) Close-up view of the foot-point (UV centroid, with $f = 0.7$, and at full cadence 24~s) trajectory and the loop-top (X-ray 6-12~keV centroid, with $f = 0.85$) trajectory between 21:44-22:00 UT. In panels (a), (b), and (d), the colors represent the times of the observed signatures as indicated by the color bar. Note that the color code in (a) and (b) is the same as in Figures~\ref{fig:hxr}, \ref{fig:reconnection}, and \ref{fig:shearevol}, but the color code in (d) is different. The curve in panels (a), (b), and (d) outlines the polarity inversion line of the radial magnetic field $B_r$. All images in this figure have been rotated to 21:00 UT, and measurements using these images reflect the coordinates at this reference time.
 }
\label{fig:motion}  
   \end{figure}

\section{Evolution of the Flare Ribbons and X-ray Loop Tops} \label{sec:motion}

The temporal and spatial evolution of flare emission signatures reflect the dynamics of reconnection energy release. Whereas the UV ribbon emission maps all energy 
release events on the chromosphere, the reconstructed X-ray sources likely only reflect the strongest events due to the limited dynamic range of the X-ray maps. In this section, we infer the total reconnection rate from the apparent motion of the UV ribbons and also measure the apparent motion of the centroids of the soft X-ray and/or UV sources to estimate the locations of prominent energy release events likely related to where non-thermal electrons are produced and deposited, respectively.

Figure~\ref{fig:motion}a shows the evolution of the newly brightened flare ribbons observed in the AIA UV 1600~\AA\ passband at a cadence of 24~s, with the color code indicating the time the ribbons start to brighten. To minimize saturation effects or transient unrelated brightenings, we detect the ribbon fronts when the brightness (in units of data counts per second) in the 1600~\AA\ passband is at least six times the pre-flare quiescent background and stays bright for at least four minutes \citep[for further discussion see][]{Naus2022}. Furthermore, we assume that the effects of reconnection only appear once at a given location so that ribbon front pixels are found only at their first brightening. %Uncertainties in the measurement are estimated by varying the threshold brightness to detect ribbon fronts, and by the difference in $\psi_+$ and $\psi_-$ measured in ribbons in positive and negative magnetic fields, respectively. (In principle, equal amounts of positive and negative fluxes should participate in reconnection.) Furthermore, the measurements presented in this study use the radial magnetic field $B_r$ rather than the longitudinal component, which leads to a smaller discrepancy between $\psi_+$ and $\psi_-$. On the other hand, we do not correct for projection effects in calculating the areas of the newly brightened ribbon fronts, and we do not extrapolate the chromospheric magnetic field (ribbons actually  form in the upper chromosphere) from the photospheric magnetic field, as these two effects partially cancel each other. These uncertainties can offset the measured total reconnection flux by up to 30\% in most flares \citep{Qiu2007}, but have lesser effects on the time evolution of the reconnection flux and the global reconnection rate.
%In a 3D configuration with component reconnection, $E_{rec}$ is not related to $\dot{\psi}$ in a simple way. In this flare,
It is evident that during the first stage, the stage of ribbon {\it elongation}, evolution of the ribbon fronts is less like what is depicted in a 2D picture where only expansion of ribbons perpendicular to the PIL would be expected. 
The ribbons spread along the PIL at an apparent speed of about 10-40 km s$^{-1}$\footnote{Following \cite{Qiu2009} and \cite {Qiu2010}, we infer the ribbon front motion with respect to the PIL. We project the positions of the ribbon fronts $d_{||}$ along the PIL and estimate the mean speed of the apparent elongation due to the extension of the ribbon along the PIL. The mean perpendicular distance of the ribbon from the PIL is estimated by $\langle d_{\perp} \rangle \approx A/l_{||}$, where $A$ is the total area of the polygon between the ribbon fronts and the section of the PIL along which the ribbon fronts are projected, and $l_{||}$ is the projected length of the ribbon fronts along the PIL.}.
The elongation parallel to the PIL halts by the end of the fast expansion stage. Both ribbons also expand in the direction perpendicular to the PIL, with a mean  speed of 3-4 km s$^{-1}$ during the {\em fast expansion} stage and then 1-2 km s$^{-1}$ in the {\em slow expansion} stage. These mean speeds in the parallel and perpendicular directions are consistent with those reported for other two-ribbon flares \citep[][and references therein]{Qiu2017}.
Note that the estimated speeds reflect the mean motion. At various locations, the ribbon front may expand much faster \citep[][and references therein]{Naus2022}.

The apparent motion of the flare ribbons is accompanied by motion of the X-ray sources shown in Figure~\ref{fig:hxr}. In particular, the X-ray emission at 6-12 keV is likely produced at or above the top of flare loops just formed by reconnection. Figure~\ref{fig:motion}b shows the trajectory of the centroid of the 6-12 keV source, $(x_{xr}, y_{xr})$, indicative of the apparent motion of the loop top. Since the cadence of the thick-target HXR maps at $\ge$ 30~keV is very low, we do not measure the centroid of these HXR sources. However, since the UV 1600~\AA\ light curve closely follows that of the non-thermal HXR emission during the impulsive phase (Figure~\ref{fig:hxr}a), we track the centroid of the UV emission as a proxy for the location of prominent thick-target non-thermal HXR emissions. Figure~\ref{fig:motion}b also displays the trajectory of the centroid $(x_{uv}, y_{uv})$ of the UV emission for the positive and negative ribbons separately (at the cadence of 72~s).%, with the size of the symbols indicating the range of $(x_{uv}, y_{uv})$ measured with different thresholds.
\footnote{The 6-12 keV X-ray centroids are measured by $x_{xr} = \sum x_i I_i/\sum I_i$ and $y_{xr} = \sum y_i I_i/\sum I_i$ with the sum conducted over pixels whose intensity $I_i$ exceeds a fraction $f$ of the peak intensity $I_m$ at that time. Varying $f$ from $0.75$ to $0.95$ does not significantly change the spatial evolution; the result for $f = 0.85$ is displayed by diamond symbols in Figure~\ref{fig:motion}b. The speed of the X-ray centroid shown in Figure~\ref{fig:motion}c is computed as the time derivative of the displacement $\Delta s = \sqrt{(\Delta x)^2 + (\Delta y)^2}$ of the centroid. The centroids of the UV emission, $(x_{uv}, y_{uv})$, are measured in the same manner with $f = 0.5, 0.7, 0.9$, and the result for $f = 0.7$ is displayed by plus symbols in Figure~\ref{fig:motion}b. Note that since the UV emission is rather extended, the UV centroid position varies significantly, by more than 10\arcsec after the impulsive phase, when measured with different thresholds.} We note that, during the impulsive phase when UV (and X-ray) emissions are significant and less dispersed, the centroid measurements with different thresholds are more consistent and therefore more reliable.

For a close look at the impulsive phase, the X-ray and UV centroids measured between 21:44-22:00 UT are further illustrated in Figure~\ref{fig:motion}d (at the cadence of 24~s). The comparison between $(x_{xr}, y_{xr})$ and $(x_{uv}, y_{uv})$ suggests a meandering motion of the sources of prominent emissions at both the chromosphere and the corona in the early phase. Up to 21:53 UT, the UV centroid in the positive magnetic field exhibits an apparent back-and-forth motion along the PIL. % with an  speed of 10-30 km s$^{-1}$.  
The X-ray 6-12 keV source moves in the same manner with a similar range of distance, %speed, 
suggesting that energy release occurs along the RCS \citep[e.g.][]{Grigis2005, Krucker2005, Inglis2013}. In addition, the UV centroid in the positive magnetic field also moves away from the PIL, with this perpendicular motion becoming faster after 21:53~UT when the 6-12 keV source speeds up as well. 
The UV centroid in the negative field exhibits similar meandering motions in the early phase, and becomes less regular later, perhaps due to the weaker and more dispersed emission on this ribbon (see Figure~\ref{fig:mmap}b). Overall, the apparent trajectory of the UV centroids suggests that the projected motion of the X-ray source at 6-12 keV may be partly due to the apparent motion along the PIL, particularly in the early phase, and partly due to the rise of the coronal source (and coincident with the separation of the two ribbons or UV centroids), which is more significant after 21:53 UT. %This will be further discussed in Section~\ref{sec:summary}.% 

The total reconnection rate, i.e., the flux change rate, can be measured from the ribbon front evolution. Figure~\ref{fig:motion}c shows $\dot{\psi}_+$ and $\dot{\psi}_-$ measured in the southern (positive) and northern (negative) ribbons, respectively.\footnote{The measurements in this study use the radial magnetic field $B_r$ rather than the longitudinal component (as used in \citet{Qiu2022}). On the other hand, we do not correct for projection effects in calculating the areas of the newly brightened ribbon fronts and we do not extrapolate the chromospheric magnetic field (ribbons actually  form in the upper chromosphere) from the photospheric magnetic field, as these two effects partially cancel each other. These uncertainties can offset the measured total reconnection flux by up to 30\% in most flares \citep{Qiu2007}, but have lesser effects on the time evolution of the reconnection flux and the global reconnection rate.} 
The two flux change rates evolve similarly and are roughly balanced. (In principle, equal amounts of positive and negative fluxes should participate in reconnection.) At the peak, $\dot{\psi}$ averaged between $\dot{\psi}_+$ and $\dot{\psi}_-$ is about 6$\times 10^{18}$ Mx s$^{-1}$. %We may estimate the average reconnection electric field by $\langle E_{rec}\rangle \approx \dot{\psi}/l_{||}$, where $l_{||}$ is the length of the ribbon extension along the PIL, estimated to be $l_{||} \approx 120$ Mm. This yields a maximum $\langle E_{rec}\rangle \approx 500$ V m$^{-1}$. We note that since the actual length of the current sheet may vary during the flare -- since reconnection can be highly non-uniform \citep{Naus2022} --  the local reconnection rate at given locations and times can be much greater than this average value. 
The apparent speed of the X-ray 6-12~keV source, $v_{top}$, is also measured and displayed in Figure~\ref{fig:motion}c. This source is accelerated during the {\em fast-expansion} stage at an average {\em projected} speed of a few tens of km s$^{-1}$, with the peak speed approaching $100$ km s$^{-1}$ between 21:53 and 21:56~UT at nearly the same time as the peak flux change rate. Observations therefore indicate the consistent evolution of the apparent motion in the corona and chromosphere, with both being indicative of reconnection dynamics. 

Significant flare emission, particularly non-thermal hard X-ray emission $\mathcal{I}_{hxr}$ above photon energies of 30~keV, occurs in the fast expansion stage. However, Figure~\ref{fig:motion}c shows that in this flare $\dot{\psi}$ and $v_{top}$ rise and peak 2-4 minutes before $\mathcal{I}_{hxr}$. In particular, the reconnection rate derived from the ribbon evolution has been much enhanced before the prominent high-energy HXR emission.  Such time lags have been reported in several prior studies that measured $\dot{\psi}$ or $\langle E_{rec} \rangle$ by tracking the ribbon fronts \citep{Falchi1997, Miklenic2007, Qiu2010, Naus2022, Vievering2023}. %It is noted that the flare HXR and total UV emissions peak at the same time (Figure~\ref{fig:hxr}a); on the other hand, the ribbon fronts are detected during the rise of the UV emission at each pixel. 
We note that many previous studies have compared $\mathcal{I}_{hxr}$ and the reconnection electric field $E_{rec}$ measured in a different way, by tracking the apparent motion of the brightest optical, UV, or HXR kernels and assuming $E_{rec} \approx v_{k} B$, $v_k$ and $B$ being the apparent motion speed of the kernel and magnetic field at the kernel. Some of these studies have revealed a temporal correlation between the two for some times and/or at some locations \citep{Qiu2002, Krucker2003, Qiu2004, Fletcher2004, Lee2006, Lee2008}, whereas others do not find a temporal or spatial correlation, particularly with refined tempo-spatial scales \citep{Grigis2005, Inglis2013}. These discrepancies suggest that the reconnection dynamics can be complicated by the field configuration, which can be more 2D-like in some flares than others. % {\bf JQ: do we understand why peak reconnection \& peak loop top motion is ahead of peak non-thermal emission by 3 minutes? This is even longer than seen in Naus2022 event.}

In a 3D reconnection configuration, the reconnection rate $E_{rec}$ is not related to $\dot{\psi}$ in a simple way since
the motion of the flare ribbons along the PIL can make a significant contribution.
Furthermore, the reconnection rate might not be the only property governing flare energetics. It has been proposed that the reconnecting guide field plays a crucial role in energizing particles \citep{wang16a,Dahlin2017, Arnold2021}. Information about the reconnection guide field may be gleaned from the observed shear of the PRFLs. In past studies, this shear angle, $\theta_{rb}$, has been inferred from observations of flare ribbons or kernels. %  $\langle d_{||}\rangle$ and $\langle d_{\perp}\rangle$ measured with the two ribbons. %This is re-plotted in Figure~\ref{fig:motion}c, showing that the inferred shear $\theta_{rb}$ varies from about 60 degree initially to less than 30 degree prior to the peak of HXR emission. 
For example, we may assume that non-thermal electrons travel to the chromosphere along ``loops'' (which may or may not exist) connecting the UV centroids in the positive and negative ribbons. Figure~\ref{fig:motion}d would then suggest that before 21:46 UT, such ``loops'' are very sheared (violet) and that in the fast expansion stage (blue, 21:46-21:52 UT), when 35-80 keV HXR is rising, the ``loops'' connecting the UV centroids become less sheared. Finally, toward the peak of the HXR emission (green, 21:52-21:58 UT), the ``loops'' are least sheared. This is consistent with the strong-to-weak shear evolution trend inferred with ribbon fronts \citep[][and references therein]{Qiu2022}. The connectivity between the centroid pair, however, is an assumption. In the next section, instead, we will employ observations of PRFLs in the EUV passbands, and make direct measurements of the shear of a large number of PRFLs, which will provide substantially more information than inferred from the evolution of flare ribbons or kernels.

\begin{figure}    %%%%%%%%%%%%%%%%%% 
\includegraphics[width=0.98\textwidth,clip=]{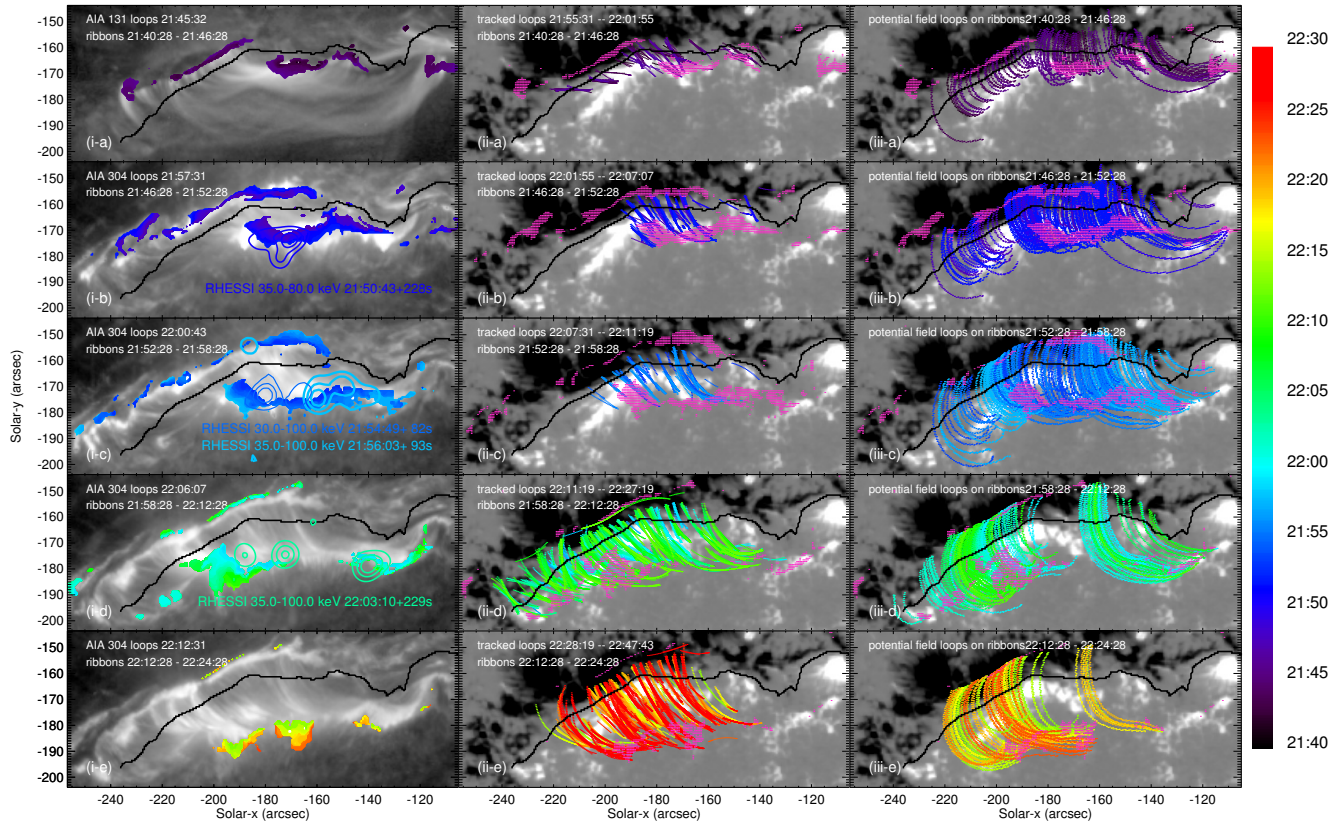}
\caption{{\em Left}: Evolution of flare ribbon fronts (color symbols) derived from the UV 1600~\AA\ images by AIA during the (a) elongation, (b-c) fast expansion, and (d-e) slow expansion phases. Superimposed are the EUV images from AIA that show post-reconnection flare loops (PRFLs) anchored at the ribbon fronts and the hard X-ray sources at $\ge$ 30~keV (color contours) obtained from RHESSI. The colors of the ribbon fronts and HXR contours indicate the times given in the color bar at right. 
{\em Middle}: PRFLs identified from AIA 304~\AA\ images, superimposed with the ribbon fronts (pink symbols) during (a) elongation, (b,c) fast-expansion, and (d,e) slow expansion, on a pre-flare magnetogram of the photospheric radial magnetic field from HMI. Colors of the PRFLs indicate the times the PRFLs are identified in the AIA 304~\AA\ images minus 15 minutes, the nominal cooling time (see text in Section \ref{subsec:cooling}), which are the same colors used in Figure~\ref{fig:shearevol}. {\em Right}: magnetic loops from the potential field extrapolation projected to the AIA image plane, superimposed on a pre-flare magnetogram of the photospheric radial magnetic field from HMI. 
%Loops are traced from ribbon fronts (pink symbols) during the different stages of the flare evolution, and colors of the loops indicate the times of the ribbon fronts shown in the left panels.
Potential field loops are traced from ribbon fronts (pink symbols) during the different stages of the flare evolution. The colors of the potential field loops (right panels) indicate the times at which the ribbon fronts formed (left panels); see color bar at right.
}  
\label{fig:reconnection}  
   \end{figure}

\begin{figure}    %%%%%%%%%%%%%%%%%% 
\includegraphics[width=0.98\textwidth,clip=]{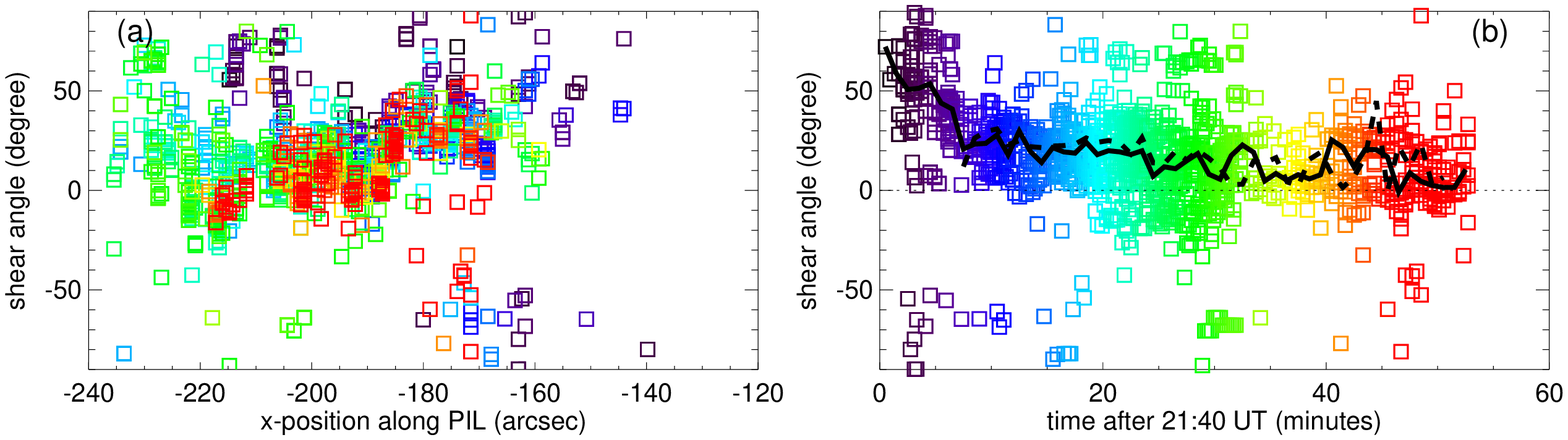}
\includegraphics[width=0.98\textwidth,clip=]{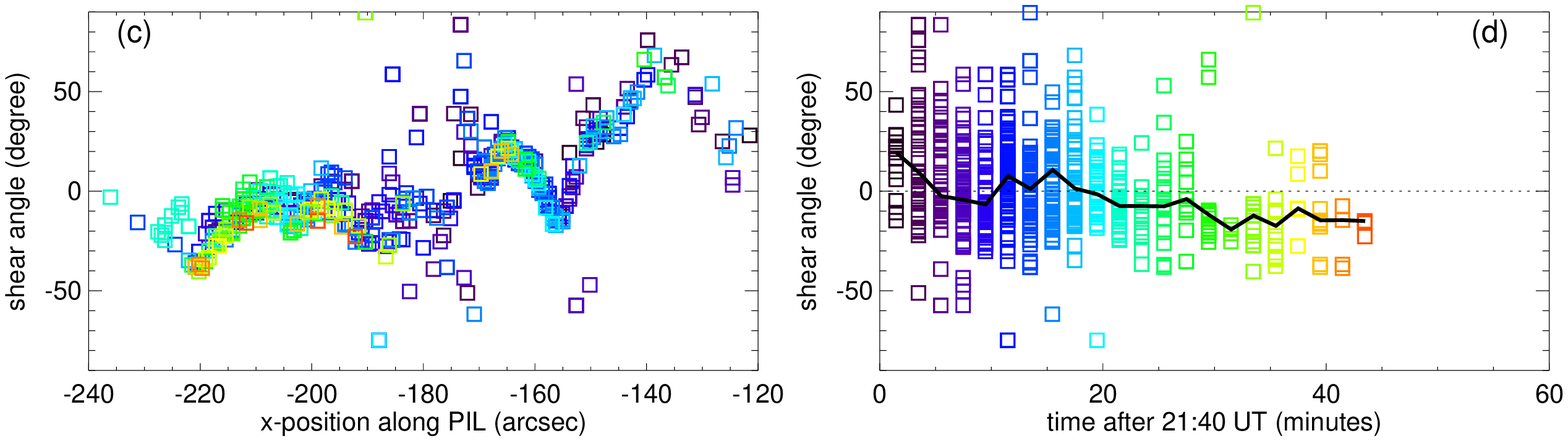}
\caption{
{\em Top}: the shear angle $\theta_{lp}$ of the PRFLs (1320 measurements) identified from AIA 304~\AA\ images with respect to the vertical of the PIL of the photosperical radial magnetic field, measured at where the PRFLs crosses the PIL, versus the $x$ position on the PIL (a) or times they are identified (b). The colors indicate the times PRFLs are identified in the AIA 304~\AA\ images minus 15 minutes (see S\ref{subsec:cooling}), which are the same as in the middle panels in Figure~\ref{fig:reconnection}. The solid black curve in (b) presents the median $\theta_{lp}$ every minute. For comparison, the dashed black curve shows the median $\theta_{lp}$ every minute of PRFLs identified in the AIA 171~\AA\ images.
{\em Bottom}: The shear angle $\theta_{pot}$  of the potential field lines (769 measurements) rooted at the flare ribbon fronts and projected to the AIA image plane with the vertical of the PIL, versus the $x$ position along the PIL (c) or the time (d) of the ribbon fronts. The color coding is the same as in the right panels of Figure~\ref{fig:reconnection}. The solid black curve shows the median $\theta_{pot}$ every two minutes.}  
\label{fig:shearevol}  
   \end{figure}

\section{Shear Evolution of Post-Reconnection Flare Loops (PRFLs)} \label{sec:shear}

As the flare evolves from the elongation to the fast expansion stage, the PRFLs become notably less sheared, exhibiting the strong-to-weak shear evolution reported in studies of many other two-ribbon flares \citep{Aschwanden2001, Ji2006, Su2006, Su2007, Liu2009b, Yang2009, Qiu2009, Qiu2010, Qiu2017}. Since the shear of the PRFLs is likely a proxy for the guide component of the magnetic field flowing into the RCS \citep{Dahlin2022}, we attempt to characterize it here. In most previous studies \citep[except][]{Qiu2017}, the shear of the PRFLs has been inferred using observations of flare ribbons or kernels exclusively. In some of these studies, the shear angle was measured between the PIL, approximated by a straight line, and another straight line connecting two dominant flare kernels in UV \citep{Su2006}, optical \citep{Ji2006}, or HXR \citep{Liu2009b, Yang2009} emissions, assuming that these are conjugate foot-points of PRFLs. The complement of this angle is defined as the shear angle $\theta$: $\theta \approx 0^{\circ}$ indicates the PRFL is perpendicular to the PIL and $\theta \approx 90^{\circ}$ refers to very high shear where the PRFL almost parallels the PIL.

In the left panels of Figure~\ref{fig:reconnection}, the strong-to-weak shear evolution of the PRFLs is apparent. However, at any given time an arcade of PRFLs is formed with their foot-points outlined by a number of flare kernels aligned along the ribbon front. Therefore, it is not directly evident from ribbon observations which pairs of kernels in opposite magnetic fields are conjugate foot-points. Furthermore, the thick-target HXR emission is mapped to a few kernels almost exclusively on one ribbon, without clear signatures of their conjugates on the other ribbon. 
%As a result, this method of estimating shear from the foot-points is not easily applicable to this flare. %without applying more complicated techniques like by \citet{Fletcher2009, Qiu2009}. 
Due to these factors, the above-described method of estimating the shear from the foot-points is not easily applied to this flare.
Instead, we will measure the shear angle $\theta$ directly using PRFLs observed in the EUV images by AIA.

\subsection{Measuring the Shear of PRFLs}\label{subsec:shear}

To do so, we first track PRFLs in the time series of EUV images. PRFLs anchored to flare ribbons formed in the elongation stage are easily visible in the EUV 131 \AA\ passband (Figure~\ref{fig:overview}e,f, Figure~\ref{fig:reconnection}i-a) and then, when these loops have cooled down sufficiently, in the EUV 304 \AA\ passband (Figure~\ref{fig:reconnection}i-b). PRFLs anchored to the flare ribbons formed later in the expansion stages are visible in the EUV 171 \AA\ passband (Figure~\ref{fig:overview}g,h; images in the 171 \AA\ passband at earlier times are saturated and not usable), as well as the EUV 304 \AA\ passband (Figure~\ref{fig:reconnection}i-b to i-e). These broadband EUV images can capture emission by plasmas in the temperature range $\le 1$~MK \citep{Odwyer2010, Boerner2012}. PRFLs visible in these passbands have cooled to these temperatures minutes after they are formed by reconnection. Therefore, the measured shear $\theta$ is delayed by their cooling time to the passband at which they are observed. We have experimented on tracking PRFLs in three passbands, in EUV 304~\AA\ images that are least subject to saturation, and in the EUV 131~\AA\ and 171~\AA\ passbands when they are not saturated before or after the peak of the flare.

To track PRFLs, we apply the algorithm of \citet{Aschwanden2010} that identifies all curvilinear structures in a given image. As unwanted byproducts, the algorithm can also pick out active region loops and, sometimes, flare ribbons. Non-PRFLs are cleaned out with a semi-automated approach guided by the geometry of ribbons. Briefly, a PRFL has to be rooted at and confined between two flare ribbons. The method is applied to more than 200 images in the 304~\AA\ passband at the full cadence (12~s per image) between 21:55 and 22:45 UT, and has successfully identified close to 2,000 PRFLs. The technique is also applied to about 90 images in the 171~\AA\ passband at half cadence (24~s per image) between 22:03 and 22:45 UT, which yields about 900 PRFLs -- images in this passband are saturated before 22:03~UT. The PRFLs found in these two different passbands are generally consistent.  The middle panels in Figure~\ref{fig:reconnection} illustrate PRFLs tracked from a series of EUV 304~\AA\ images, superimposed on the $B_r$ map and the ribbon fronts (pink symbols) during different stages of the flare evolution, where the color code indicates the times the PRFLs are observed (minus 15 minutes, the nominal cooling time of the PRFLs; see Section~\ref{subsec:cooling} for more discussion). 

Qualitatively, it is evident from Figure~\ref{fig:reconnection} that PRFLs in the early stage are more sheared, i.e., more inclined toward the PIL, than those later. Strictly speaking, the shear of a PRFL is a 3D property that is not feasible to determine without a realistic model of the magnetic configuration of the reconnection current sheet. As an alternative method we compare the geometry of the observed PRFLs with the extrapolated potential field lines projected to the AIA image plane. These are traced from all 5,800 ribbon front pixels in both the positive and negative magnetic polarities. The right panels of Figure~\ref{fig:reconnection} show a subset of potential field lines anchored at ribbon fronts; those field lines traced from northern ribbons (negative field) and southern ribbons (positive field), respectively. Colors indicate the time when the ribbon pixels are brightened. The comparison of the observed PRFLs with the potential field indicates that PRFLs deviate more from the potential field in the early phases of the flare, i.e., the elongation phase and early expansion phase. 

Such a comparison can be quantified by measuring the angle made by a PRFL (or a potential field line projected in the AIA image plane) with the PIL at where it crosses the PIL. By convention, this angle ranges between 0 and 180 degrees, measured clockwise from the east (the PIL roughly follows the east-west direction). We define the complement of this angle as the shear angle, $\theta_{lp}$ for PRFLs and $\theta_{pot}$ for the potential field. The shear angle $\theta_{lp}$ is measured for all PRFLs, yielding more than 1,300 valid measurements (i.e., when the PRFL crosses the PIL). The angle $\theta_{pot}$ is measured in one-fifth of all 5,800 potential field lines projected to the AIA image plane, yielding more than 700 valid measurements. Figure~\ref{fig:shearevol}a-b shows the measured $\theta_{lp}$ for about 1,300 PRFLs identified in the AIA 304~\AA\ images, along the PIL (panel a) during the flare evolution (panel b). Colors indicate the times the PRFLs are observed (minus 15 minutes) and are the same as in the middle panels of Figure~\ref{fig:reconnection}. Initially $\theta_{lp}$ is as high as 60-70$^{\circ}$, but over a period of 10 minutes, its median decreases to about 20$^{\circ}$ and then continues to decrease gradually as the flare evolves. In comparison, the shear of the potential field $\theta_{pot}$ also exhibits a decreasing trend, but its median starts at 20$^{\circ}$ and then decreases to around $0 \pm 10$$^{\circ}$. We note the difference in the spatial distributions of the potential field loops and the observed PRFLs. For example, Figure~\ref{fig:shearevol}a and \ref{fig:shearevol}c show that the observed PRFLs extend to the east of $-$210\arcsec, during the slow expansion stage, whereas there are a larger number of potential field loops west of $-$160\arcsec. However, a comparison of the shear evolution of a subset of modeled and observed loops crossing the PIL only between $-$210\arcsec\ and $-$160\arcsec\ finds that the trend of the shear evolution of the subsets is not changed significantly. This analysis supports the strong-to-weak shear evolution of PRFLs, which is also consistent with the trend inferred qualitatively from the apparent motion of the ribbon fronts or UV centroids.

\begin{figure}    %%%%%%%%%%%%%%%%%% 
\includegraphics[width=0.9\textwidth,clip=]{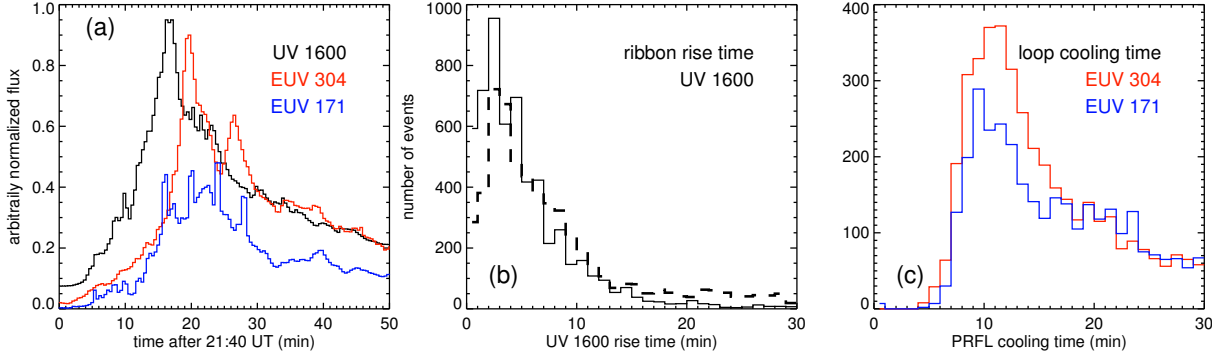}
\caption{
(a) The light curve of total UV 1600 \AA\ emission from the flare ribbons (black), in comparison with that of the total emission of the EUV 304~\AA\ (red) or the EUV 171~\AA\ (blue) at locations along the PIL. (b) Histograms of the rise times of UV 1600 \AA\ emission at 5,800 ribbon front pixels, as either the time it takes for the UV emission to rise from six times the pre-flare quiescent brightness to its peak (solid), or as the width of the half-Gaussian used to approximate the UV light curve from its rise to peak (dashed). (c) Histograms of the cooling times of PRFLs to the EUV 304~\AA\ and 171~\AA\ passbands estimated with the UFC model.}  
\label{fig:cooling}  
   \end{figure}

\subsection{Cooling Times of PRFLs}\label{subsec:cooling}

The potential field is traced from the locations of the ribbon fronts, which are brightened at the times PRFLs are just formed by reconnection. The PRFLs then cool
down to the necessary $\le$1~MK to produce prominent emissions in the 304~\AA\ (or 171 \AA) passband.  We can estimate this cooling time in several ways. First, Figure~\ref{fig:cooling}a compares the light curve of the total UV 1600 \AA\ emission $\mathcal{I}_{1600}$ from ribbons with that of the total EUV 304 \AA\ emission $\mathcal{I}_{304}$ at the locations along the PIL that sample loops connecting the two ribbons. The peaks of $\mathcal{I}_{304}$ lag those of $\mathcal{I}_{1600}$ by $\sim$5~min. Figure~\ref{fig:cooling}b shows the rise time $\tau_{rise}$ of the UV emission at each ribbon pixel as either the time it takes for the UV emission to rise from six times the pre-flare quiescent brightness to its peak, or as the width of the half-Gaussian used to approximate the UV light curve from its rise to peak. Either way, the statistical analysis shows that, in this flare, $\tau_{rise}$ of the UV 1600~\AA\ emission in the majority ($\ge$ 70\%) of 5,800 ribbon-front pixels is larger than 2 minutes, with the median $\tau_{rise}$ being 4-5~min. Taking this rise time into account, it takes about 10 minutes, on average, for reconnection-formed PRFLs to produce prominent emission in the EUV 304~\AA\ passband. 

Next, we estimate the PRFL cooling times using the Ultraviolet Foot-point Calorimeter (UFC) method to model evolution of the flare loops with heating rates inferred from the foot-point UV light curves \citep{Qiu2012, Zhu2018, Qiu2021}. As a first-order estimate, the lengths of these loops are computed using the potential field extrapolation. This way, the 5,800 (half-)loops, assumed to be anchored at 5,800 ribbon pixels, are modeled and the synthetic total X-ray and EUV emissions from these loops are compared against observations by GOES and AIA, which allows constraints to be placed on the few free parameters used in the model. Once reasonable agreement between the observed and synthetic total X-ray/EUV emissions has been achieved, we obtain the synthetic time profiles of the EUV emission in the AIA 304~\AA\ passband from individual loops (again, assumed to be anchored at the ribbon pixels) and estimate the time lags $\tau_{304}$ of the peak EUV emission in these loops with respect to the times when their feet are brightened in the UV 1600~\AA\ passband. Figure~\ref{fig:cooling}c shows histograms of the cooling times $\tau_{304}$ and $\tau_{171}$. Statistically, $\tau_{304}$ is found to lie between 5 and 30 minutes, with the mode at 11 minutes and median at 16 minutes. The time lags can also be estimated as the difference between the peak UV 1600 \AA\ emission at the ribbon front pixel and the peak synthetic EUV 304 \AA\ emission in the (half-)loop anchored to the foot-point.  The mode and median of these lags are 4 minutes and 9 minutes respectively -- recall that the median of the rise time, $\tau_{rise}$, of the foot-point UV emission is 5 minutes.  The time lags of the loop emission in 171~\AA\ are similar, suggesting that PRFLs shown in these two passbands emit at similar temperatures. These time lags, or the cooling times of PRFLs, are shown to grow with their length -- shorter loops cool more quickly than longer loops. Estimated with these different approaches, the cooling time of the bulk of the observed PRFLs in the 304~\AA\ and 171~\AA\ passbands ranges between 5 and 15 minutes. Neglecting such variations, we take $\langle \tau_{304} \rangle \approx 15$ minutes for all the PRFLs as a nominal cooling time, and 
shift the times of the PRFLs backward by 15 minutes in the middle panels in Figure~\ref{fig:reconnection} and Figure~\ref{fig:shearevol}a-b. 

Finally, we compare the variation of shear $\theta_{lp}$ measured from 1,300 PRFLs observed in the EUV 304~\AA\ passband with the shear $\theta_{rb}$ that is inferred from the mean positions of the ribbon fronts \citep[see details of the method in][]{Qiu2022}.\footnote{\citet{Qiu2022} measured the shear index $\mathcal{S}$, which is equivalent to the tangent of the shear angle $\theta_{rb}$ if the PIL is assumed or approximated to be a straight line. Also note that $\theta_{rb} \approx \tan^{-1}(\mathcal{S})$ is a crude measurement of the shear of a ``loop" assumed to connect the average position of the ribbon fronts in the positive field and that in the negative field.} The brightening of the ribbon fronts essentially coincides with the time PRFLs are just formed by reconnection. Figure~\ref{fig:spectrum}a shows the median of $\theta_{lp}$ over every minute (red; shifted back by 15 minutes) in comparison with $\theta_{rb}$ (orange). The two independent measurements show consistent strong-to-weak shear evolution, but over different time scales. This is due to the varying cooling times of the PRFLs to the EUV 304~\AA\ passband. Specifically, in the early or impulsive phase of the flare, $\tau_{304}$ is expected to be shorter, of order 5-10 minutes, than in the late phase.

Although it is difficult to establish a one-to-one association between PRFLs observed in EUV images and their foot-points observed in the UV 1600 \AA\ images, the comparison of the observed shear of PRFLs and that of the potential field loops anchored at flare ribbon fronts provides quantitative evidence supporting the strong-to-weak shear evolution of PRFLs. A more accurate, one-to-one comparison can be achieved with improved magnetic and hydrodynamic models of the PRFLs, which will be pursued in future work.

\section{Flare Energetics and Reconnection Properties} \label{sec:HXR}
\begin{figure}    %%%%%%%%%%%%%%%%%% 
        %\centerline{\includegraphics[width=0.5\textwidth,clip=]{fig6_v2.pdf}}
%        \includegraphics[width=0.50\textwidth,clip=]{fig8a.pdf}
\begin{interactive}{animation}{movie_20141218.mp4}        
        \includegraphics[width=0.47\textwidth,clip=]{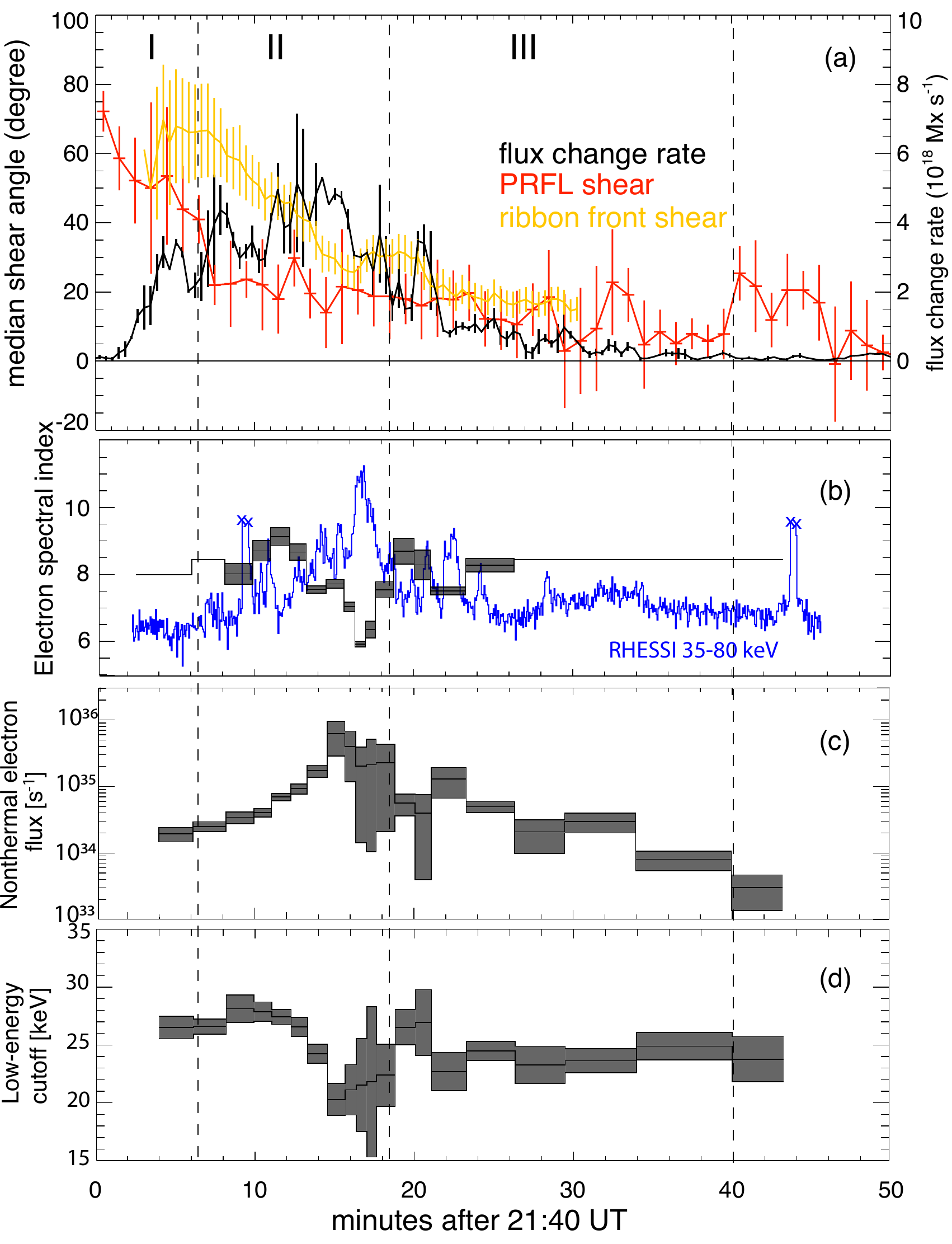}
\includegraphics[width=0.46\textwidth,clip=]{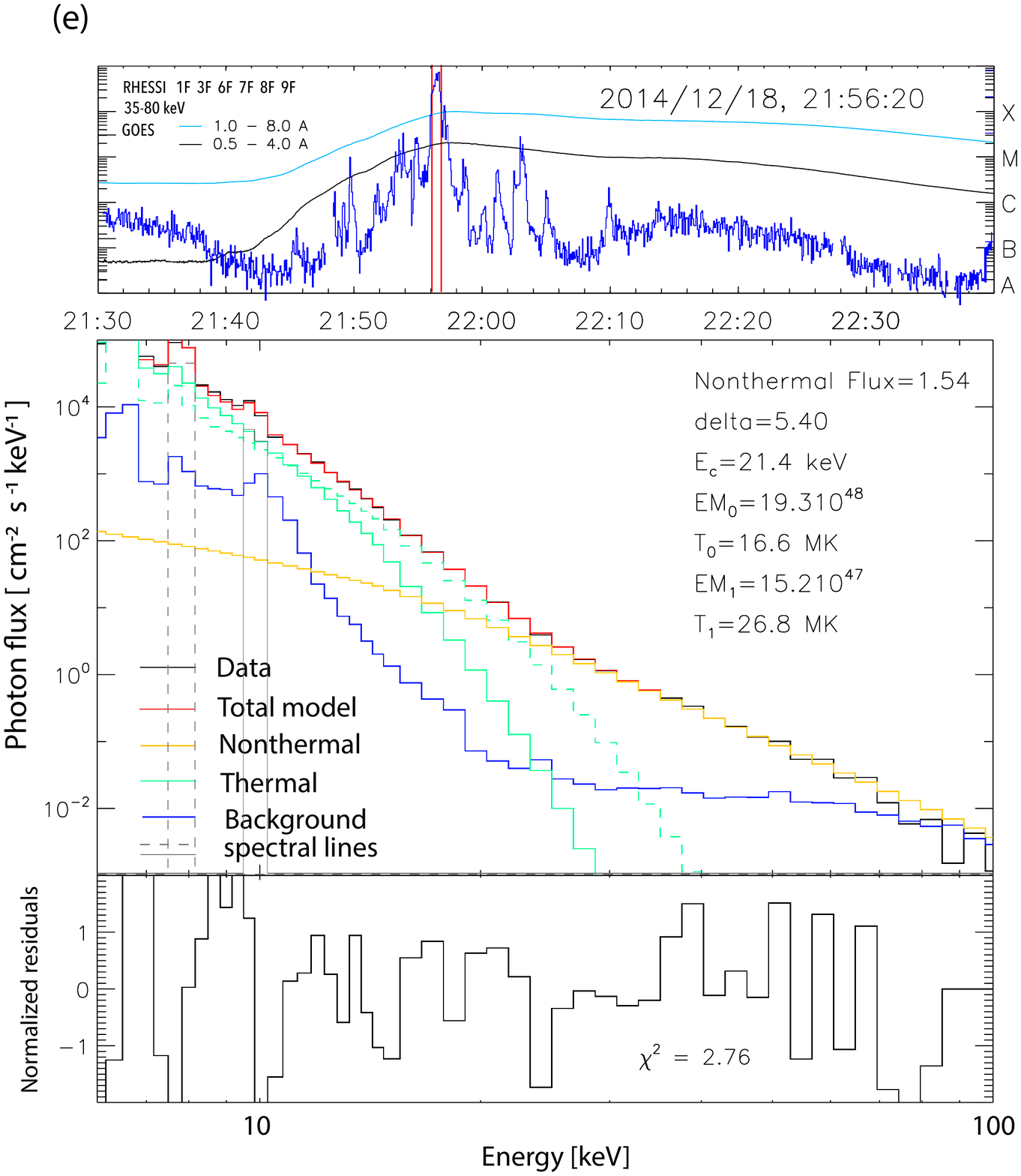}
\end{interactive}
            \caption{Flare parameters versus time.  (a): The flux change rate (black, 24~s cadence), median PRFL shear ($\theta_{lp}$, red, 1-min cadence ), and ribbon front shear ($\theta_{rb}$, orange,  24~s cadence). The displayed flux change rate is the average of $\dot{\psi}_+$ and $\dot{\psi}_-$, with vertical bars indicating the range of the rate measured in the positive and negative fields. The red vertical bars in the $\theta_{lp}$ plot indicate the one-half of the standard deviation of the measured $\theta_{lp}$ every minute. The orange vertical bars in the $\theta_{rb}$ plot show the standard deviation of the measurements using varying thresholds to identify ribbon fronts \citep{Qiu2022}.
            (b): The HXR 35-80 keV flux (blue). (b-d): The non-thermal electron distribution parameters (black) derived from fitting the hard X-ray spectra, including the electron spectral index, the total non-thermal electron flux, and the low-energy cutoff, respectively. The width of the curves represents the $1\sigma$ uncertainty on the respective fit parameters. Fit time intervals are non-uniform. (e): An example of the fit to the X-ray spectrum at the peak of the HXR emission, showing the X-ray light curves by GOES and RHESSI and the time interval of the fit (top), the observed spectrum and the best fit to it with fitting parameters (middle), and the normalized residuals of the fit and the reduced $\chi^2$ (bottom). The fits to the spectra throughout the flare are displayed in the attached supplemental movie.  The animation lasts 21~s and shows the results of the RHESSI spectral fits similarly to the right panel. The complete figure set (19 images) is available in the online journal. Each image represents the time interval of successive spectral fits.} 

               \label{fig:spectrum}  
   \end{figure}

To understand the implication of the shear on flare energetics, we compare its evolution measured with the PRFLs against other properties. Figure~\ref{fig:spectrum}a shows the median of $\theta_{lp}$ over every minute (red), as well as $\theta_{rb}$ inferred from the ribbon fronts (orange) and the flux change rate $\dot{\psi}$ (both at the cadence of 24~s). The light curve of the HXR 35-80~keV counts is given in Figure~\ref{fig:spectrum}b.
As discussed in Section~\ref{sec:motion}, $\dot{\psi}$ rises and peaks ahead of the HXR emission; meanwhile, $\langle \theta_{lp}\rangle$ or $\theta_{rb}$ starts high and decreases, during which time the observed $\ge 30$~keV HXR emission rises toward its peak. The flare HXR emission is a proxy for the flux carried by the non-thermal electrons, and the shear is a proxy of the relative guide field in the RCS.  %\citep{Dahlin2022}. 
%In the following, we derive properties of the non-thermal electrons from HXR spectral analysis and explore how their flux may be related to the guide field obtained from the observationally measured shear. 
In this section, we derive properties of the non-thermal electrons from HXR spectral analysis and relate them to the observationally measured shear.

We conduct spectroscopic analysis of the flare X-ray emissions observed by RHESSI to derive properties of non-thermal electrons.\footnote{HXR fluxes were detected by RHESSI up to $\sim100$~keV. A spatially-integrated spectral analysis using detector 6 was performed using a model with two thermal components, a single power law consistent with the collisional thick-target model, two physical spectral lines at 6.7~keV and 8~keV and an instrumental line around 10 keV that is needed to obtain a good spectral fit. Analyses using detectors 1 and 3 separately give similar results. In addition, the spectra were corrected for pulse pile-up and albedo effects assuming an isotropic distribution of electrons (the default parameters in OSPEX). The fitting procedure used here is different from that in \citet{Qiu2022}, who only fitted the photon spectrum, not the electron spectrum, and used an iso-thermal model with one thermal component plus a broken power-law.}  Panels (b-d) of Figure~\ref{fig:spectrum} show the non-thermal electron distribution parameters: the electron spectral index, the total non-thermal electron flux $\mathcal{F}_e$, and the low-energy cutoff, respectively. The widths of the curves represent the $1\sigma$ uncertainty on the respective fit parameters. At time intervals before 21:48 and after 22:06, the electron spectral index is fixed at the plotted values. These were adjusted to provide the best fits and to reduce the number of free parameters because the counts above 30~keV are significantly reduced. Therefore, the value of the spectral index is plotted but no uncertainty is provided. Figure~\ref{fig:spectrum}e shows an example of the fit to the observed spectrum at the peak of the HXR emission. The fits to all the spectra, integrated with varying intervals depending on the counts, are provided in the supplemental movie.

The flux of accelerated electrons is coupled to the value of the low-energy cutoff. The spectral fits reveal a flattening at lower energies, i.e., below 30-50 keV, during the HXR peak time between 21:56  and 21:58 UT, whereas we deduce a single power-law without any significant flattening before and after this interval. While the electron spectral index at higher energies reflects that of the accelerated distribution, the spectral index at lower energies can have other sources, including propagation effects, for example, through deceleration of the non-thermal beam by the co-spatial return current electric field \citep[e.g.,][and references therein]{2006ApJ...651..553Z,2020ApJ...902...16A,2021ApJ...917...74A} or non-uniform target ionization \citep[][]{su2011}, or even  instrumental effects \citep[see][for reviews on the low-energy cutoff and mechanisms affecting the HXR spectra]{holman2011,kontar2011}.

The accelerated distribution can include a double power law that would appear either as a gradually flattening spectrum toward lower energies or a low-energy cutoff value higher than the transition energy between the thermal and non-thermal portions of the X-ray spectrum. However, the interpretation adopted in this paper is an HXR spectrum that flattens as a consequence of a low-energy cutoff \citep[e.g.,][]{holman2003}. Although it is known that a sharp low-energy cutoff is unstable to wave-particle interactions \citep[e.g.,][]{emslie2003,hannah2009}, its adoption is customary, both to simplify calculations of the non-thermal electron flux and because it is usually indistinguishable from a gradually flattening low-energy cutoff \citep[][]{psh2005}. During time intervals where the HXR spectrum is consistent with a single power-law (without a flattening at lower energies), only the maximum low-energy cutoff and minimum electron flux can be deduced from the spectra. This corresponds to all the intervals before and after the peak of the impulsive phase at 21:56:20-21:58:00. Conversely, as the cutoff is needed to explain the flattening, \textit{under the assumption of an injected single power-law electron distribution in the collisional thick target model}, at the above-mentioned HXR peak times the value of the electron flux (and low-energy cutoff) is determined rather than its lower limit (and upper limit, respectively). Note that the total non-thermal flux of electrons peaks ahead of the hard X-ray emission at 35-80~keV, possibly because of the higher deduced value of the low-energy cutoff during the HXR peak, similarly to \citet{2009ApJ...699..917W}. 

The peak of the magnetic flux change rate $\dot{\psi}$ in Figure \ref{fig:spectrum}a is nearly coincident with the peak of the non-thermal electron flux $\mathcal{F}_e$ in Figure~\ref{fig:spectrum}c. On the other hand, $\dot{\psi}$ has already been enhanced in the first 10 minutes of the flare, when the non-thermal flux is insignificant. 
%This will be explored in more detail in section~\ref{sec:shear_nth}. 
This relationship will be explored in the following section.

\section{Inferences from Modeling}\label{sec:model}

Neither the magnetic field nor the energy distribution of electrons in the reconnection current sheet of a solar flare can be measured directly. However, recent advances in theory and numerical modeling of eruptive flares and reconnecting current sheets may provide significant insights into the observed evolution of eruptive events, such as the flare studied in detail here. Even the most sophisticated such simulations are, of necessity, far simpler than any actual event occurring in nature. Nevertheless, just as the principles of the canonical CSHKP model provide basic understanding of our observations, more recent investigations extend and deepen this understanding in important ways. The connections between the reconnection guide field and the PRFL shear on the one hand, and the guide field and nonthermal electron acceleration on the other, are explored below.

%\subsection{Inferring the Relative Guide Field}\label{subsec:model}

\subsection{Relative Guide Field}\label{subsec:guidefield}

The shear measured from PRFLs or inferred from the ribbons is a proxy of the relative guide field at the RCS. 
%, which is not directly observable. 
This quantity is not directly observable. However, detailed numerical models of the 3D reconnection configuration can be exploited to infer the relative reconnection guide field from the measured shear.
%Numerical models of the 3D reconnection configuration can be explored to translate the measured shear to the guide field. 
For this purpose, we present results from a high-resolution three-dimensional magnetohydrodynamic calculation of an eruptive flare, described in detail by \citet{Dahlin2022}. This simulation was performed with the Adaptively Refined Magnetohydrodynamics Solver
\citep[ARMS;][]{DeVore2008} and employed an idealized magnetic configuration consisting of two sets of dipoles located just beneath the solar surface at the equator, forming an elongated polarity inversion line aligned with the equator. Shear flux was injected at this PIL using the STITCH method \citep[STatistical InjecTion of Condensed Helicity;][and references therein]{Dahlin2022b} to form a filament channel that eventually erupted via the breakout mechanism \citep{Antiochos1999}.

\begin{figure}
\includegraphics[width=0.96\textwidth,clip=]{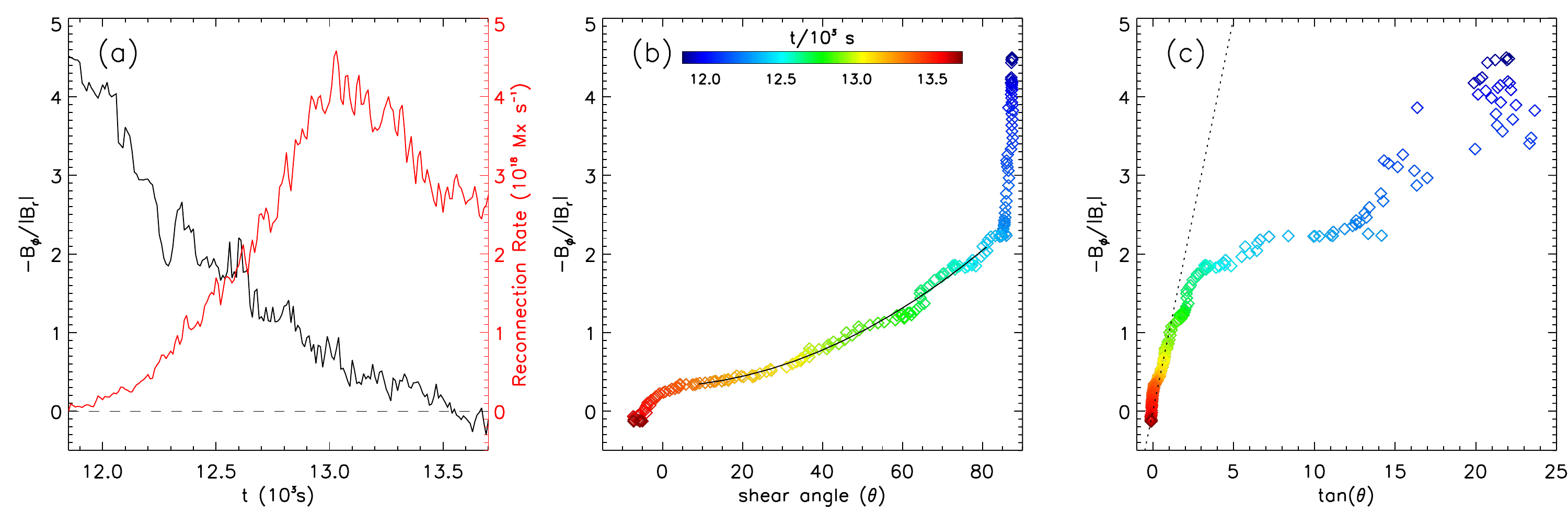}
	\caption{Guide field and shear angle evolution in the ARMS eruptive flare model. (a) Guide-field ratio ($-B_\phi/B_r$) calculated upstream of the reconnecting current sheet at $\phi=0$ (black) and rate of total reconnected flux (red). Guide field ratio versus the mean shear angle ($\theta$) is shown in (b) and versus its tangent in (c). The mean shear angle is the angle between the foot-points of a newly reconnected flare loop and the direction normal to the PIL, averaged over the region $|\phi| < 2^\circ$. The guide field is calculated at $\phi = 0$, and the upstream is taken to be the location where the current density first attains 25\% of its peak value when approaching the current sheet. The color indicates the time when the newly reconnected flare loops are identified and the corresponding guide field is calculated.  The solid line in the center panel is a parabolic fit and the dashed line in the right panel corresponds to $-B_{\phi}/|B_r| = \tan(\theta).$}
\label{fig:shearbg}  
   \end{figure}

To investigate the relationship between reconnection properties at the RCS and observables, namely PRFLs and ribbons,  %of the magnetic shear to the guide field, 
we traced field lines from a grid of $901\times226$ foot-points at the inner boundary of the simulation. Our criterion for identifying reconnection events was a shortening of the field-line length by $40\%$ relative to its maximum value. We then measured the reconnection flux $\psi$ underlying these foot-points of shortening field lines and computed the reconnection rate $\dot{\psi}$. We also estimated the ratio of the guide field (the $B_{\phi}$ or longitudinal component in our simulation coordinates) to the reconnected field (the $B_r$ or radial component) upstream of the current sheet at zero longitude (the center of the configuration). 
%The time evolutions of $\dot{\psi}$ and the relative guide field $\mathcal{R} \equiv B_{\phi}/B_r$ are plotted in Figure~\ref{fig:shearbg}a, showing that the guide field ratio $\mathcal{R} \gtrsim 0.5$ before the reconnection rate peaks and $\mathcal{R} \lesssim 0.5$ afterward, similar to what is observed in the M6.9 flare. 
The time evolutions of $\dot{\psi}$ and the relative guide field $\mathcal{R} \equiv B_{\phi}/B_r$ are plotted in Figure~\ref{fig:shearbg}a, showing that the guide field ratio $\mathcal{R} \gtrsim 0.75$ before the reconnection rate peaks and $\mathcal{R} \lesssim 0.75$ afterward.
We then calculated the shear angles $\theta$ from the conjugate foot-points of the resulting flare loops, and 
%present a `lookup table' (Fig.~\ref{fig:shearbg}) to 
generated figures that 
relate the guide field to the PRFL shear. The mean shear angle (averaged over the region $|\phi| < 2^\circ$) is plotted against the guide field ratio at $10$ s cadence. A parabolic curve fit for the range $9^\circ < \theta < 81^\circ$ is shown in Figure~\ref{fig:shearbg}b. 
At a guide-field ratio of 0.75, the mean shear angle is about 35$^\circ$. 
For comparison, the observed M6.9 flare had an average PRFL shear of about 20$^\circ$ at the peak of the flare (Figure \ref{fig:spectrum}a). 
This corresponds to a guide-field ratio of about 0.40 in the simulation (Figure \ref{fig:shearbg}b). 
Finally, 
Figure~\ref{fig:shearbg}c shows that for $|B_{\phi}/B_r|\lesssim 2$ (or $\theta \lesssim 60^{\circ}$) the scaling $|B_{\phi}/B_r| = \tan{\theta}$ holds. 
%Note that these relations are derived from a model describing a symmetric configuration with a straight PIL and two ribbons parallel to the PIL, and therefore may provide a reference for flares, such as the M6.9 flare studied here, with a similarly simple geometry.

We emphasize that the relations above are derived from a model describing a symmetric configuration with a straight PIL and two ribbons parallel to that PIL. Detailed quantitative agreement with any particular observed flare cannot be expected. Nevertheless, the results provide a baseline reference for flares that have a relatively simple geometry, such as the M6.9 flare studied in this paper. Specifically, we find that the values (0.40, 0.75) of the guide-field ratio $\mathcal{R}$ at the times of peak flux change rate agree within a factor of two. Both values are consistent with a guide field that is somewhat weaker than the reconnecting field components in the RCS. The observed HXR flux peaks slightly later than the flux change rate, when the guide-field ratio is steady or slowly decreasing further. This finding is consistent with recent models for electron acceleration in reconnecting current sheets, as discussed below.

%In Fig.~\ref{fig:bgplot}, we show the same guide field ratio as plotted in Fig.~\ref{fig:shearbg} against time, compared to the rate of reconnected flux, which shows that the guide field ratio $b_g \gtrsim 0.5$ before the reconnection peaks and $b_g \lesssim 0.5$ afterward, suggesting that peak particle acceleration would be delayed with respect to the peak reconnection rate, as seen in the observations.

\subsection{Magnetic Shear and Non-Thermal Electron Production}\label{sec:shear_nth}
\begin{figure}[ht!]
 \includegraphics[width=0.95\textwidth]{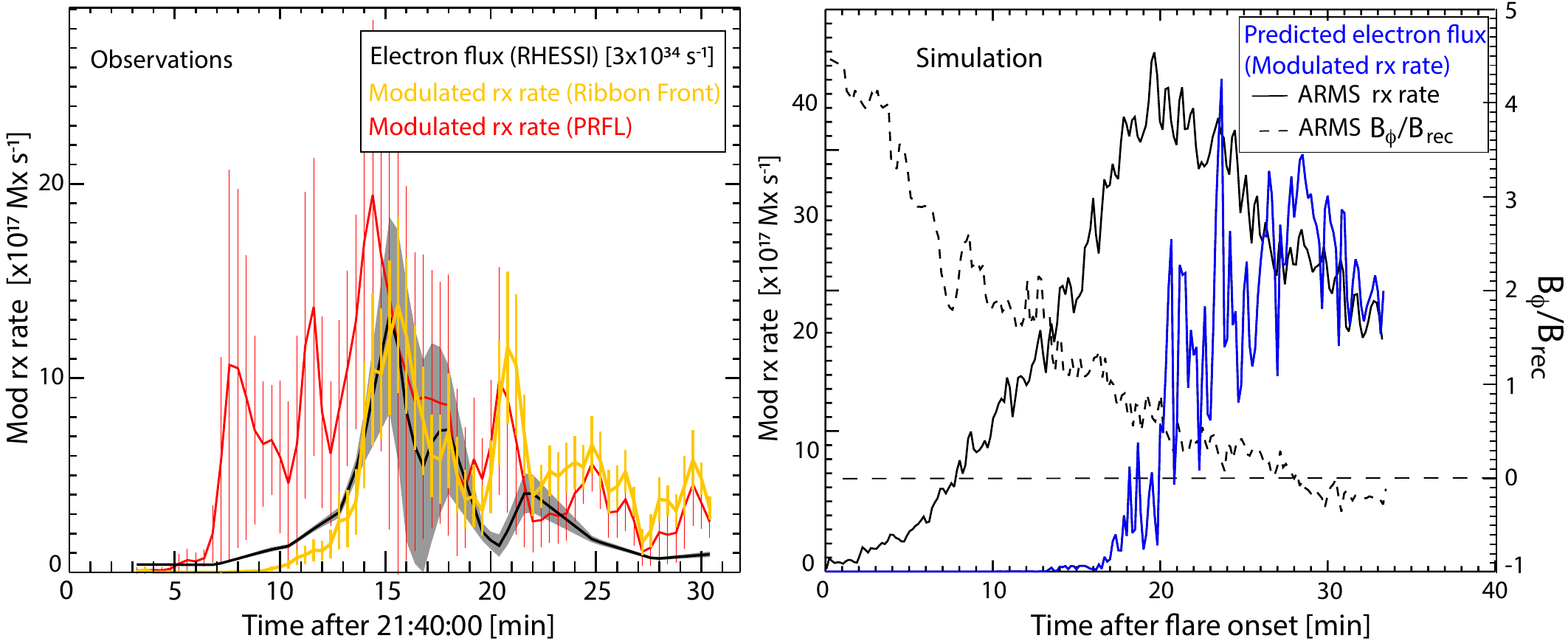}
\caption{{\em Left:} Temporal evolution of the non-thermal electron flux $\mathcal{F}_e$ as deduced from RHESSI spectral fits 
(black curve with gray uncertainties), and the modulated magnetic flux change rate using the deduced shear from ribbon fronts 
(orange) and the shear deduced from the PRFLs (red), which are shifted backward by a nominal cooling time of 15 minutes. 
{\em Right:} the modulated reconnection rate $\dot{\psi}_{mod}$ (blue), or the model predicted non-thermal electron flux, 
calculated from Equation \ref{psimod}, using the reconnection rate $\dot{\psi}$ (solid black) and the relative guide field $\mathcal{R}$ 
(dashed black) from the numerical simulation shown in Figure \ref{fig:shearbg}a. }
 \label{fig:mod_rx_rate_electrons}
\end{figure}

Theoretical models of non-thermal electron production during magnetic reconnection suggest that a dominant control parameter is the magnetic shear upstream of the reconnection current sheet. %\citep{Arnold2021}. 
An empirical fit to the results of the numerical simulations of \citet{Arnold2021} -- see their Figure 4c --  finds that the fraction of non-thermal electrons $f_{nt}=n_{nt}/(n_{nt}+n_t)$ scales as $\text{sech}^2(2.4B_{g}/B_{rec})$, where $B_{g}$ is the guide field and is related to the shear by $B_{g}/B_{rec} \approx \tan{\theta}$ (see Figure \ref{fig:shearbg}c of this paper).  
A rough scaling law for the rate of production of non-thermal electrons then follows by multiplying the total number of electrons injected into the current layer by this fraction
\begin{equation}\label{nt}
\dot{n}_{nt} \approx f_{nt}n_{tot}V_rL^2 \approx f_{nt}n_{tot}\dot{\psi}L/B_{rec},
\end{equation}
in which $V_r$ is the characteristic reconnection inflow speed and $L$ is the characteristic scale length of the flare current sheet. Thus, the modulation of the non-thermal electron production rate as the guide field changes during a flare can be written as $\dot{n}_{nt}=\dot{\psi}_{mod} n_{tot} L/B_{rec},$ where the shear-modulated
total reconnection rate is given by
\begin{equation}\label{psimod}
\dot{\psi}_{mod}=\dot{\psi}\,\text{sech}^2\left(2.4\frac{B_g}{B_{rec}}\right).
\end{equation}
The above equation suggests that, although other parameters $n_{tot}$, $L$, and $B_{rec}$ may vary during the flare, the modulation due to the changing relative guide field has the largest impact on the production of non-thermal electrons. Figure~\ref{fig:mod_rx_rate_electrons}a combines the electron flux $\mathcal{F}_{e}$ determined from the RHESSI spectral fits, the modulated magnetic flux change rate calculated from equation \ref{psimod}, using the magnetic shear calculated in both ways discussed above.  The modulated reconnection rate using $\theta_{rb}$ has a similar time history to the electron fluxes up to the peak and for about 5 minutes following.  Although the correlation diminishes after that point, so do the calculated RHESSI electron fluxes, suggesting that these times do not contribute significantly to the total non-thermal electron production. The modulated reconnection rate using $\langle \theta_{lp} \rangle$ shifted back by a nominal cooling time of 15 minutes is not as well correlated. Nevertheless, there are uncertainties in the cooling times of the PRFLs, and $\tau_{304}$ of PRFLs formed in the impulsive phase are expected to be shorter than 15 min (see section~\ref{subsec:cooling}), which would bring $\dot{\psi}_{mod}$ closer to $\mathcal{F}_{e}$. Pursuit of an improved estimate of the time evolution of $\dot{\psi}_{mod}$ -- perhaps with improved estimates of cooling times of observed PRFLs that will also permit the establishment of the spatial distribution of the shear with respect to energetic electrons -- will be left to future work.  In Fig.~\ref{fig:mod_rx_rate_electrons}b, we show $\dot{\psi}_{mod}$ computed with $\dot{\psi}$ and $B_{\phi}/B_{r}$ from the model (Figure~\ref{fig:shearbg}a)  suggesting that peak particle acceleration would be delayed with respect to the peak reconnection rate, as seen in the observations.

\section{Summary and Conclusions}\label{sec:summary}

We have analyzed the evolution of an M6.9 two-ribbon flare to study the properties of the triggering reconnection as well as the flare-accelerated non-thermal electrons.
For the first time, the shear of post-reconnection flare loops has been measured using several independent techniques enabling a %direct 
cross validation of the obtained estimates. The results obtained by these complementary techniques are in a reasonable quantitative and an excellent qualitative agreement. Observational measurements of this M6.9 flare lead to the following findings. %and they all demonstrate the following main tendencies during  the post-reconnection evolution of the studied ribbon system.} 

\begin{itemize}
\itemsep0em 
\setlength{\itemsep}{0pt}
\item An enhanced reconnection rate leads prominent flare emissions, particularly the thick-target non-thermal HXR emission, by several minutes. 
%The flare also exhibits strong-to-weak shear evolution, with the PRFL shear varying from 60-70 degrees prior to the HXR onset to below 20 degrees during the impulsive rise of the non-thermal HXR. 
\item The median shear of PRFLs decreases monotonically during the impulsive phase. 
\item The non-thermal electron flux $\mathcal{F}_{e}$ peaks when $\dot{\psi}$ is nearly maximal and the median shear of the PRFLs satisfies $\langle \theta_{lp} \rangle \approx 20^{\circ}$. 
%\item The presence of the large shear of the PRFLs may be indicative of a large guide field component $B_g$, relative to the reconnection component $B_{rec}$,  along the invariant direction in the RCS and the time variation of the shear is related to the change of the ratio $\mathcal{R} \equiv B_g/B_{rec}$ as reconnection progresses.  
%\item A large shear angle of the PRFLs is indicative of a large guide-field component $B_g$, relative to the reconnection component $B_{rec}$, along the invariant direction in the RCS.
%\item The temporal variation of the shear is related to the change of the ratio $\mathcal{R} \equiv B_g/B_{rec}$ as the reconnection progresses.
\item An MHD model of an eruptive flare confirms that the temporal variation of the shear is related to the change of the ratio $\mathcal{R} \equiv B_g/B_{rec}$ in the RCS.
\item Models of electron acceleration in a reconnecting current sheet indicate that acceleration becomes more efficient for $\mathcal{R} \lesssim 1$ \citep{Dahlin2017,Arnold2021}.
\item The observations and models are fully consistent with the HXR fluxes peaking later than the reconnection rate and, in this particular case at least, long after the initial onset of flare reconnection.
%(Since it is difficult to model the 3D structure of the RCS in the real event, we use more idealized MHD models to help infer the relative guide field  $\mathcal{R}$ from the observationally measured shear angle even though apparent motions of the flare ribbons and the X-ray loop top emission indicate flare reconnection beyond the standard 2D geometry.) 
\end{itemize}

Our results confirm the strong-to-weak shear evolution reported in previous observational and numerical studies. The analysis shows that during the first ten minutes of forming strongly sheared PRFLs in this flare, magnetic reconnection is not an efficient producer of energetic non-thermal electrons. Similarly, energetic electrons are not prevalent in the late phase when the shear of the PFRLs is near zero yet the reconnection rate is low. These results suggest that intermediate shear is needed, $\theta \le 40^\circ$, for efficient particle acceleration via reconnection. 

Past observational studies \citep[][and references therein]{Qiu2022} have inferred the evolution of the magnetic shear of PRFLs by tracking the foot-points or ribbons with the assumed connectivities between one or a few pairs of foot-points. This study takes advantage of the AIA observations of a multitude of PRFLs, and directly measures the angles made by the (projected) PRFLs with the PIL at where they cross. The high-cadence (12~s) and continuous AIA observations in multiple passbands make it possible to derive more than one thousand $\theta_{lp}$ measurements, which is a substantial progress, in both quality and quantity, over the $\theta_{rb}$ measurements. The comparison between $\theta_{rb}$ and the mean $\theta_{lp}$, for this specific event, shows that measurements in different passbands and with two independent methods are consistent, thus validating the practice to infer the shear by tracking the evolution of flare ribbons or foot-points. This study also demonstrates that PRFLs are not potential (Section~\ref{sec:shear}). In future work, three-dimensional magnetic structure of PRFLs may be reconstructed guided by the projected PRFLs identified from observations, and the spatial distribution of the magnetic shear and of flare radiation signatures (UV, EUV, and HXR) will be compared. These efforts will advance our understanding of three-dimensional magnetic reconnection and energy release. Such experiment will also be expanded to more flares to test the general validity of the methods and conclusions based on this event.

The PRFL shear is considered to be a proxy of the relative guide field $\mathcal{R} \equiv B_g/B_{rec}$ in the current sheet. The observed shear evolution is indicative of the reconnection configuration and dynamics, and the phenomenological relation with the non-thermal emission suggests that the reconnection guide field plays a crucial role in flare energetics. 
This role can be further clarified in future studies combining sophisticated data analysis techniques with data-constrained numerical simulations. The physical explanation of the nonlinear relation between the shear angle and the non-thermal electron production proposed in our paper has been tested by idealized MHD and PIC models. In future investigations, it will be important to model the 3D structure of the RCS for real flaring events in order to infer $\mathcal{R}$ from the observationally measured shear angles. Such data-constrained 3D modeling would also help resolve ambiguities associated with apparent motions of the flare ribbons and the X-ray loop top emission indicating flare reconnection beyond the 2D geometry. 

%% Also note that the akcnowlodgment environment does not support long amounts of text. If you have a lot of people and institutions to acknowledge, do not use this command. Instead, create a new \section{Acknowledgments}.

\begin{acknowledgments}

We thank the referee for constructive comments that have helped improve the clarity of the manuscript. We thank Drs. Judy Karpen and Dale Gary for 
discussions. The collaboration leading to these results was facilitated by the NASA Drive Science Center on Solar Flare Energy
Release (SolFER), Grant No.\ 80NSSC20K0627. Resources supporting this work were provided by the NASA High-End Computing (HEC) Program through the NASA Center for Climate Simulation (NCCS) at Goddard Space Flight Center. J.Q.\ was supported by NASA grants Nos.\ 80NSSC22K0519 and 80NSSC23K0414. M.A. was supported by NASA grants 80NSSC23K0043 and 80NSSC20K1813. S.K.A.\ was supported by the Partnership for Heliophysics and Space Environmental Research between U MD and NASA/GSFC. J.T.D.\ was supported by NASA grants Nos.\ 80NSSC21K1313 and 80NSSC21K0816. C.R.D.\ was supported by NASA's H-ISFM program at GSFC. J.F.D.\ and M.S.\ were also supported by NSF Grants Nos.\ PHY1805829 and PHY2109083 and NASA Grant No.\ 80NSSC20K1813.  A.R. was supported by the NSF's REU program at Montana State University. V.M.U.\ was partly supported through the Partnership for Heliophysics and Space Environment Research (NASA grant No.\ 80NSSC21M0180).
{\em SDO} is a mission of NASA's Living With a Star Program. The authors also thank Kim Tolbert for technical support and the {\em RHESSI} Mission Archive for the data support. % Plus, I think the previous sentence is from the template and should eventually be deleted along with this comment.} {\jtdfont JD: The NCCS acknowledgment is for the ARMS simulation.}

\end{acknowledgments}

%Authors may use the online length %calculator to get an estimate of 
%the number of word and float quanta in %their manuscript. The calculator 
%is located at %\url{https://authortools.aas.org/Quanta/%newlatexwordcount.html}.


\bibliography{shear}{}

\begin{thebibliography}{}
\expandafter\ifx\csname natexlab\endcsname\relax\def\natexlab#1{#1}\fi
\providecommand{\url}[1]{\href{#1}{#1}}
\providecommand{\dodoi}[1]{doi:~\href{http://doi.org/#1}{\nolinkurl{#1}}}
\providecommand{\doeprint}[1]{\href{http://ascl.net/#1}{\nolinkurl{http://ascl.net/#1}}}
\providecommand{\doarXiv}[1]{\href{https://arxiv.org/abs/#1}{\nolinkurl{https://arxiv.org/abs/#1}}}

\bibitem[{{Alaoui} {et~al.}(2021){Alaoui}, {Holman}, {Allred}, \&
  {Eufrasio}}]{2021ApJ...917...74A}
{Alaoui}, M., {Holman}, G.~D., {Allred}, J.~C., \& {Eufrasio}, R.~T. 2021,
  \apj, 917, 74, \dodoi{10.3847/1538-4357/ac0820}

\bibitem[{{Allred} {et~al.}(2020){Allred}, {Alaoui}, {Kowalski}, \&
  {Kerr}}]{2020ApJ...902...16A}
{Allred}, J.~C., {Alaoui}, M., {Kowalski}, A.~F., \& {Kerr}, G.~S. 2020, \apj,
  902, 16, \dodoi{10.3847/1538-4357/abb239}

\bibitem[{{Antiochos} {et~al.}(1999){Antiochos}, {DeVore}, \&
  {Klimchuk}}]{Antiochos1999}
{Antiochos}, S.~K., {DeVore}, C.~R., \& {Klimchuk}, J.~A. 1999, \apj, 510, 485,
  \dodoi{10.1086/306563}

\bibitem[{{Arnold} {et~al.}(2021){Arnold}, {Drake}, {Swisdak}, {Guo}, {Dahlin},
  {Chen}, {Fleishman}, {Glesener}, {Kontar}, {Phan}, \& {Shen}}]{Arnold2021}
{Arnold}, H., {Drake}, J.~F., {Swisdak}, M., {et~al.} 2021, \prl, 126, 135101,
  \dodoi{10.1103/PhysRevLett.126.135101}

\bibitem[{{Aschwanden}(2010)}]{Aschwanden2010}
{Aschwanden}, M.~J. 2010, \solphys, 262, 399, \dodoi{10.1007/s11207-010-9531-6}

\bibitem[{{Aschwanden} \& {Alexander}(2001)}]{Aschwanden2001}
{Aschwanden}, M.~J., \& {Alexander}, D. 2001, \solphys, 204, 91,
  \dodoi{10.1023/A:1014257826116}

\bibitem[{{Aschwanden} {et~al.}(2019){Aschwanden}, {Kontar}, \&
  {Jeffrey}}]{Aschwanden2019}
{Aschwanden}, M.~J., {Kontar}, E.~P., \& {Jeffrey}, N. L.~S. 2019, \apj, 881,
  1, \dodoi{10.3847/1538-4357/ab2cd4}

\bibitem[{{Boerner} {et~al.}(2012){Boerner}, {Edwards}, {Lemen}, {Rausch},
  {Schrijver}, {Shine}, {Shing}, {Stern}, {Tarbell}, {Title}, {Wolfson},
  {Soufli}, {Spiller}, {Gullikson}, {McKenzie}, {Windt}, {Golub}, {Podgorski},
  {Testa}, \& {Weber}}]{Boerner2012}
{Boerner}, P., {Edwards}, C., {Lemen}, J., {et~al.} 2012, \solphys, 275, 41,
  \dodoi{10.1007/s11207-011-9804-8}

\bibitem[{{Bogachev} {et~al.}(2005){Bogachev}, {Somov}, {Kosugi}, \&
  {Sakao}}]{Bogachev2005}
{Bogachev}, S.~A., {Somov}, B.~V., {Kosugi}, T., \& {Sakao}, T. 2005, \apj,
  630, 561, \dodoi{10.1086/431918}

\bibitem[{{Carmichael}(1964)}]{Carmichael1964}
{Carmichael}, H. 1964, NASSP, 50, 451

\bibitem[{{Caspi} {et~al.}(2014){Caspi}, {Krucker}, \& {Lin}}]{Caspi2014}
{Caspi}, A., {Krucker}, S., \& {Lin}, R.~P. 2014, \apj, 781, 43,
  \dodoi{10.1088/0004-637X/781/1/43}

\bibitem[{{Cheng} {et~al.}(2012){Cheng}, {Kerr}, \& {Qiu}}]{Cheng2012}
{Cheng}, J.~X., {Kerr}, G., \& {Qiu}, J. 2012, \apj, 744, 48,
  \dodoi{10.1088/0004-637X/744/1/48}

\bibitem[{{Dahlin}(2020)}]{Dahlin2020}
{Dahlin}, J.~T. 2020, PhPl, 27, 100601, \dodoi{10.1063/5.0019338}

\bibitem[{{Dahlin} {et~al.}(2022{\natexlab{a}}){Dahlin}, {Antiochos}, {Qiu}, \&
  {DeVore}}]{Dahlin2022}
{Dahlin}, J.~T., {Antiochos}, S.~K., {Qiu}, J., \& {DeVore}, C.~R.
  2022{\natexlab{a}}, \apj, 932, 94, \dodoi{10.3847/1538-4357/ac6e3d}

\bibitem[{{Dahlin} {et~al.}(2022{\natexlab{b}}){Dahlin}, {DeVore}, \&
  {Antiochos}}]{Dahlin2022b}
{Dahlin}, J.~T., {DeVore}, C.~R., \& {Antiochos}, S.~K. 2022{\natexlab{b}},
  \apj, 941, 79, \dodoi{10.3847/1538-4357/ac9e5a}

\bibitem[{{Dahlin} {et~al.}(2015){Dahlin}, {Drake}, \& {Swisdak}}]{Dahlin2015}
{Dahlin}, J.~T., {Drake}, J.~F., \& {Swisdak}, M. 2015, PhPl, 22, 100704,
  \dodoi{10.1063/1.4933212}

\bibitem[{Dahlin {et~al.}(2017)Dahlin, Drake, \& Swisdak}]{Dahlin2017}
Dahlin, J.~T., Drake, J.~F., \& Swisdak, M. 2017, PhPl, 24,
  \dodoi{10.1063/1.4986211}

\bibitem[{{Daou} \& {Alexander}(2016)}]{Daou2016}
{Daou}, A.~G., \& {Alexander}, D. 2016, \apj, 832, 63,
  \dodoi{10.3847/0004-637X/832/1/63}

\bibitem[{{Dennis} \& {Pernak}(2009)}]{2009ApJ...698.2131D}
{Dennis}, B.~R., \& {Pernak}, R.~L. 2009, \apj, 698, 2131,
  \dodoi{10.1088/0004-637X/698/2/2131}

\bibitem[{{DeVore} \& {Antiochos}(2008)}]{DeVore2008}
{DeVore}, C.~R., \& {Antiochos}, S.~K. 2008, \apj, 680, 740,
  \dodoi{10.1086/588011}

\bibitem[{{Emslie}(2003)}]{emslie2003}
{Emslie}, A.~G. 2003, \apjl, 595, L119, \dodoi{10.1086/378168}

\bibitem[{{Emslie} {et~al.}(2012){Emslie}, {Dennis}, {Shih}, {Chamberlin},
  {Mewaldt}, {Moore}, {Share}, {Vourlidas}, \& {Welsch}}]{Emslie2012}
{Emslie}, A.~G., {Dennis}, B.~R., {Shih}, A.~Y., {et~al.} 2012, \apj, 759, 71,
  \dodoi{10.1088/0004-637X/759/1/71}

\bibitem[{{Falchi} {et~al.}(1997){Falchi}, {Qiu}, \& {Cauzzi}}]{Falchi1997}
{Falchi}, A., {Qiu}, J., \& {Cauzzi}, G. 1997, \aap, 328, 371

\bibitem[{{Fletcher} \& {Hudson}(2001)}]{Fletcher2001}
{Fletcher}, L., \& {Hudson}, H. 2001, \solphys, 204, 69,
  \dodoi{10.1023/A:1014275821318}

\bibitem[{{Fletcher} {et~al.}(2004){Fletcher}, {Pollock}, \&
  {Potts}}]{Fletcher2004}
{Fletcher}, L., {Pollock}, J.~A., \& {Potts}, H.~E. 2004, \solphys, 222, 279,
  \dodoi{10.1023/B:SOLA.0000043580.89730.4d}

\bibitem[{{Forbes} \& {Lin}(2000)}]{Forbes2000}
{Forbes}, T.~G., \& {Lin}, J. 2000, JASTP, 62, 1499,
  \dodoi{10.1016/S1364-6826(00)00083-3}

\bibitem[{{Forbes} \& {Priest}(1984)}]{Forbes1984}
{Forbes}, T.~G., \& {Priest}, E.~R. 1984, in Solar Terrestrial Physics: Present
  and Future (NASA), ed. D.~M. {Butler} \& K.~{Papadopoulos}, 1--35

\bibitem[{{Grigis} \& {Benz}(2005)}]{Grigis2005}
{Grigis}, P.~C., \& {Benz}, A.~O. 2005, \apjl, 625, L143,
  \dodoi{10.1086/431147}

\bibitem[{{Hannah} {et~al.}(2009){Hannah}, {Kontar}, \& {Sirenko}}]{hannah2009}
{Hannah}, I.~G., {Kontar}, E.~P., \& {Sirenko}, O.~K. 2009, \apjl, 707, L45,
  \dodoi{10.1088/0004-637X/707/1/L4510.48550/arXiv.0911.0314}

\bibitem[{{Hinterreiter} {et~al.}(2018){Hinterreiter}, {Veronig}, {Thalmann},
  {Tschernitz}, \& {P{\"o}tzi}}]{Hinterreiter2018}
{Hinterreiter}, J., {Veronig}, A.~M., {Thalmann}, J.~K., {Tschernitz}, J., \&
  {P{\"o}tzi}, W. 2018, \solphys, 293, 38, \dodoi{10.1007/s11207-018-1253-1}

\bibitem[{{Hirayama}(1974)}]{Hirayama1974}
{Hirayama}, T. 1974, \solphys, 34, 323, \dodoi{10.1007/BF00153671}

\bibitem[{{Holman}(2003)}]{holman2003}
{Holman}, G.~D. 2003, \apj, 586, 606, \dodoi{10.1086/367554}

\bibitem[{{Holman} {et~al.}(2011){Holman}, {Aschwanden}, {Aurass}, {Battaglia},
  {Grigis}, {Kontar}, {Liu}, {Saint-Hilaire}, \& {Zharkova}}]{holman2011}
{Holman}, G.~D., {Aschwanden}, M.~J., {Aurass}, H., {et~al.} 2011, \ssr, 159,
  107, \dodoi{10.1007/s11214-010-9680-910.48550/arXiv.1109.6496}

\bibitem[{{Inglis} \& {Gilbert}(2013)}]{Inglis2013}
{Inglis}, A.~R., \& {Gilbert}, H.~R. 2013, \apj, 777, 30,
  \dodoi{10.1088/0004-637X/777/1/30}

\bibitem[{{Isobe} {et~al.}(2002){Isobe}, {Shibata}, \& {Machida}}]{Isobe2002}
{Isobe}, H., {Shibata}, K., \& {Machida}, S. 2002, GeoRL, 29, 2014,
  \dodoi{10.1029/2001GL013816}

\bibitem[{{Isobe} {et~al.}(2005){Isobe}, {Takasaki}, \& {Shibata}}]{Isobe2005}
{Isobe}, H., {Takasaki}, H., \& {Shibata}, K. 2005, \apj, 632, 1184,
  \dodoi{10.1086/444490}

\bibitem[{{Ji} {et~al.}(2006){Ji}, {Huang}, {Wang}, {Zhou}, {Li}, {Zhang}, \&
  {Song}}]{Ji2006}
{Ji}, H., {Huang}, G., {Wang}, H., {et~al.} 2006, \apjl, 636, L173,
  \dodoi{10.1086/500203}

\bibitem[{{Jing} {et~al.}(2007){Jing}, {Lee}, {Liu}, {Gary}, \&
  {Wang}}]{Jing2007}
{Jing}, J., {Lee}, J., {Liu}, C., {Gary}, D.~E., \& {Wang}, H. 2007, \apjl,
  664, L127, \dodoi{10.1086/520812}

\bibitem[{{Jing} {et~al.}(2005){Jing}, {Qiu}, {Lin}, {Qu}, {Xu}, \&
  {Wang}}]{Jing2005}
{Jing}, J., {Qiu}, J., {Lin}, J., {et~al.} 2005, \apj, 620, 1085,
  \dodoi{10.1086/427165}

\bibitem[{{Joshi} {et~al.}(2017){Joshi}, {Sterling}, {Moore}, {Magara}, \&
  {Moon}}]{Joshi2017}
{Joshi}, N.~C., {Sterling}, A.~C., {Moore}, R.~L., {Magara}, T., \& {Moon},
  Y.-J. 2017, \apj, 845, 26, \dodoi{10.3847/1538-4357/aa7c1b}

\bibitem[{{Kawaguchi} {et~al.}(1982){Kawaguchi}, {Kurokawa}, {Funakoshi}, \&
  {Nakai}}]{Kawaguchi1982}
{Kawaguchi}, I., {Kurokawa}, H., {Funakoshi}, Y., \& {Nakai}, Y. 1982,
  \solphys, 78, 101, \dodoi{10.1007/BF00151146}

\bibitem[{{Kazachenko} {et~al.}(2017){Kazachenko}, {Lynch}, {Welsch}, \&
  {Sun}}]{Kazachenko2017}
{Kazachenko}, M.~D., {Lynch}, B.~J., {Welsch}, B.~T., \& {Sun}, X. 2017, \apj,
  845, 49, \dodoi{10.3847/1538-4357/aa7ed6}

\bibitem[{{Kitahara} \& {Kurokawa}(1990)}]{Kitahara1990}
{Kitahara}, T., \& {Kurokawa}, H. 1990, \solphys, 125, 321,
  \dodoi{10.1007/BF00158409}

\bibitem[{{Kontar} {et~al.}(2011){Kontar}, {Brown}, {Emslie}, {Hajdas},
  {Holman}, {Hurford}, {Ka{\v{s}}parov{\'a}}, {Mallik}, {Massone}, {McConnell},
  {Piana}, {Prato}, {Schmahl}, \& {Suarez-Garcia}}]{kontar2011}
{Kontar}, E.~P., {Brown}, J.~C., {Emslie}, A.~G., {et~al.} 2011, \ssr, 159,
  301, \dodoi{10.1007/s11214-011-9804-x}

\bibitem[{{Kopp} \& {Pneuman}(1976)}]{Kopp1976}
{Kopp}, R.~A., \& {Pneuman}, G.~W. 1976, \solphys, 50, 85,
  \dodoi{10.1007/BF00206193}

\bibitem[{{Krucker} {et~al.}(2005){Krucker}, {Fivian}, \& {Lin}}]{Krucker2005}
{Krucker}, S., {Fivian}, M.~D., \& {Lin}, R.~P. 2005, AdSpR, 35, 1707,
  \dodoi{10.1016/j.asr.2005.05.054}

\bibitem[{{Krucker} {et~al.}(2011){Krucker}, {Hudson}, {Jeffrey}, {Battaglia},
  {Kontar}, {Benz}, {Csillaghy}, \& {Lin}}]{Krucker2011}
{Krucker}, S., {Hudson}, H.~S., {Jeffrey}, N.~L.~S., {et~al.} 2011, \apj, 739,
  96, \dodoi{10.1088/0004-637X/739/2/96}

\bibitem[{{Krucker} {et~al.}(2003){Krucker}, {Hurford}, \& {Lin}}]{Krucker2003}
{Krucker}, S., {Hurford}, G.~J., \& {Lin}, R.~P. 2003, \apjl, 595, L103,
  \dodoi{10.1086/378840}

\bibitem[{{Lee} \& {Gary}(2008)}]{Lee2008}
{Lee}, J., \& {Gary}, D.~E. 2008, \apjl, 685, L87, \dodoi{10.1086/592292}

\bibitem[{{Lee} {et~al.}(2006){Lee}, {Gary}, \& {Choe}}]{Lee2006}
{Lee}, J., {Gary}, D.~E., \& {Choe}, G.~S. 2006, \apj, 647, 638,
  \dodoi{10.1086/505416}

\bibitem[{{Liu} \& {Wang}(2009)}]{Liu2009a}
{Liu}, C., \& {Wang}, H. 2009, \apjl, 696, L27,
  \dodoi{10.1088/0004-637X/696/1/L27}

\bibitem[{{Liu} {et~al.}(2009){Liu}, {Petrosian}, {Dennis}, \&
  {Holman}}]{Liu2009b}
{Liu}, W., {Petrosian}, V., {Dennis}, B.~R., \& {Holman}, G.~D. 2009, \apj,
  693, 847, \dodoi{10.1088/0004-637X/693/1/847}

\bibitem[{{Melrose} \& {White}(1981)}]{Melrose1981}
{Melrose}, D.~B., \& {White}, S.~M. 1981, JGRA, 86, 2183,
  \dodoi{10.1029/JA086iA04p02183}

\bibitem[{{Miklenic} {et~al.}(2007){Miklenic}, {Veronig}, {Vr{\v{s}}nak}, \&
  {Hanslmeier}}]{Miklenic2007}
{Miklenic}, C.~H., {Veronig}, A.~M., {Vr{\v{s}}nak}, B., \& {Hanslmeier}, A.
  2007, \aap, 461, 697, \dodoi{10.1051/0004-6361:20065751}

\bibitem[{{Moore} {et~al.}(2001){Moore}, {Sterling}, {Hudson}, \&
  {Lemen}}]{Moore2001}
{Moore}, R.~L., {Sterling}, A.~C., {Hudson}, H.~S., \& {Lemen}, J.~R. 2001,
  \apj, 552, 833, \dodoi{10.1086/320559}

\bibitem[{{Naus} {et~al.}(2022){Naus}, {Qiu}, {DeVore}, {Antiochos}, {Dahlin},
  {Drake}, \& {Swisdak}}]{Naus2022}
{Naus}, S.~J., {Qiu}, J., {DeVore}, C.~R., {et~al.} 2022, \apj, 926, 218,
  \dodoi{10.3847/1538-4357/ac4028}

\bibitem[{{O'Dwyer} {et~al.}(2010){O'Dwyer}, {Del Zanna}, {Mason}, {Weber}, \&
  {Tripathi}}]{Odwyer2010}
{O'Dwyer}, B., {Del Zanna}, G., {Mason}, H.~E., {Weber}, M.~A., \& {Tripathi},
  D. 2010, \aap, 521, A21, \dodoi{10.1051/0004-6361/201014872}

\bibitem[{{Patsourakos} {et~al.}(2020){Patsourakos}, {Vourlidas},
  {T{\"o}r{\"o}k}, {Kliem}, {Antiochos}, {Archontis}, {Aulanier}, {Cheng},
  {Chintzoglou}, {Georgoulis}, {Green}, {Leake}, {Moore}, {Nindos}, {Syntelis},
  {Yardley}, {Yurchyshyn}, \& {Zhang}}]{Patsourakos20}
{Patsourakos}, S., {Vourlidas}, A., {T{\"o}r{\"o}k}, T., {et~al.} 2020, \ssr,
  216, 131, \dodoi{10.1007/s11214-020-00757-9}

\bibitem[{{Poletto} \& {Kopp}(1986)}]{Poletto1986}
{Poletto}, G., \& {Kopp}, R.~A. 1986, in The lower atmosphere of solar flares,
  ed. D.~F. {Neidig} \& M.~E. {Machado} (Sunspot NM: National Solar
  Observatory), 453

\bibitem[{{Qiu}(2009)}]{Qiu2009}
{Qiu}, J. 2009, \apj, 692, 1110, \dodoi{10.1088/0004-637X/692/2/1110}

\bibitem[{{Qiu}(2021)}]{Qiu2021}
---. 2021, \apj, 909, 99, \dodoi{10.3847/1538-4357/abe0b3}

\bibitem[{{Qiu} \& {Cheng}(2022)}]{Qiu2022}
{Qiu}, J., \& {Cheng}, J. 2022, \solphys, 297, 80,
  \dodoi{10.1007/s11207-022-02003-7}

\bibitem[{{Qiu} {et~al.}(2007){Qiu}, {Hu}, {Howard}, \& {Yurchyshyn}}]{Qiu2007}
{Qiu}, J., {Hu}, Q., {Howard}, T.~A., \& {Yurchyshyn}, V.~B. 2007, \apj, 659,
  758, \dodoi{10.1086/512060}

\bibitem[{{Qiu} {et~al.}(2002){Qiu}, {Lee}, {Gary}, \& {Wang}}]{Qiu2002}
{Qiu}, J., {Lee}, J., {Gary}, D.~E., \& {Wang}, H. 2002, \apj, 565, 1335,
  \dodoi{10.1086/324706}

\bibitem[{{Qiu} {et~al.}(2010){Qiu}, {Liu}, {Hill}, \& {Kazachenko}}]{Qiu2010}
{Qiu}, J., {Liu}, W., {Hill}, N., \& {Kazachenko}, M. 2010, \apj, 725, 319,
  \dodoi{10.1088/0004-637X/725/1/319}

\bibitem[{{Qiu} {et~al.}(2012){Qiu}, {Liu}, \& {Longcope}}]{Qiu2012}
{Qiu}, J., {Liu}, W.-J., \& {Longcope}, D.~W. 2012, \apj, 752, 124,
  \dodoi{10.1088/0004-637X/752/2/124}

\bibitem[{{Qiu} {et~al.}(2017){Qiu}, {Longcope}, {Cassak}, \&
  {Priest}}]{Qiu2017}
{Qiu}, J., {Longcope}, D.~W., {Cassak}, P.~A., \& {Priest}, E.~R. 2017, \apj,
  838, 17, \dodoi{10.3847/1538-4357/aa6341}

\bibitem[{{Qiu} {et~al.}(2004){Qiu}, {Wang}, {Cheng}, \& {Gary}}]{Qiu2004}
{Qiu}, J., {Wang}, H., {Cheng}, C.~Z., \& {Gary}, D.~E. 2004, \apj, 604, 900,
  \dodoi{10.1086/382122}

\bibitem[{{Saba} {et~al.}(2006){Saba}, {Gaeng}, \& {Tarbell}}]{Saba2006}
{Saba}, J.~L.~R., {Gaeng}, T., \& {Tarbell}, T.~D. 2006, \apj, 641, 1197,
  \dodoi{10.1086/500631}

\bibitem[{{Saint-Hilaire} \& {Benz}(2005)}]{psh2005}
{Saint-Hilaire}, P., \& {Benz}, A.~O. 2005, \aap, 435, 743,
  \dodoi{10.1051/0004-6361:20041918}

\bibitem[{{Sakao}(1994)}]{Sakao1994}
{Sakao}, T. 1994, PhD thesis, Univ.\ Tokyo

\bibitem[{{Schou} {et~al.}(2012){Schou}, {Scherrer}, {Bush}, {Wachter},
  {Couvidat}, {Rabello-Soares}, {Bogart}, {Hoeksema}, {Liu}, {Duvall}, {Akin},
  {Allard}, {Miles}, {Rairden}, {Shine}, {Tarbell}, {Title}, {Wolfson},
  {Elmore}, {Norton}, \& {Tomczyk}}]{Schou2012}
{Schou}, J., {Scherrer}, P.~H., {Bush}, R.~I., {et~al.} 2012, \solphys, 275,
  229, \dodoi{10.1007/s11207-011-9842-2}

\bibitem[{{Sturrock}(1966)}]{Sturrock1966}
{Sturrock}, P.~A. 1966, \nat, 211, 695, \dodoi{10.1038/211695a0}

\bibitem[{{Su} {et~al.}(2007){Su}, {Golub}, \& {Van Ballegooijen}}]{Su2007}
{Su}, Y., {Golub}, L., \& {Van Ballegooijen}, A.~A. 2007, \apj, 655, 606,
  \dodoi{10.1086/510065}

\bibitem[{{Su} {et~al.}(2011){Su}, {Holman}, \& {Dennis}}]{su2011}
{Su}, Y., {Holman}, G.~D., \& {Dennis}, B.~R. 2011, \apj, 731, 106,
  \dodoi{10.1088/0004-637X/731/2/106}

\bibitem[{{Su} {et~al.}(2006){Su}, {Golub}, {van Ballegooijen}, \&
  {Gros}}]{Su2006}
{Su}, Y.~N., {Golub}, L., {van Ballegooijen}, A.~A., \& {Gros}, M. 2006,
  \solphys, 236, 325, \dodoi{10.1007/s11207-006-0039-z}

\bibitem[{{Svestka}(1980)}]{Svestka1980}
{Svestka}, Z. 1980, RSPTA, 297, 575, \dodoi{10.1098/rsta.1980.0233}

\bibitem[{{Temmer} {et~al.}(2007){Temmer}, {Veronig}, {Vr{\v s}nak}, \&
  {Miklenic}}]{Temmer2007}
{Temmer}, M., {Veronig}, A.~M., {Vr{\v s}nak}, B., \& {Miklenic}, C. 2007,
  \apj, 654, 665, \dodoi{10.1086/509634}

\bibitem[{{Vievering} {et~al.}(2023){Vievering}, {Vourlidas}, {Zhu}, {Qiu}, \&
  {Glesener}}]{Vievering2023}
{Vievering}, J.~T., {Vourlidas}, A., {Zhu}, C., {Qiu}, J., \& {Glesener}, L.
  2023, \apj, 946, 81, \dodoi{10.3847/1538-4357/acbe3d}

\bibitem[{{Vorpahl}(1976)}]{Vorpahl1976}
{Vorpahl}, J.~A. 1976, \apj, 205, 868, \dodoi{10.1086/154343}

\bibitem[{Wang {et~al.}(2016)Wang, Lu, Huang, \& Wang}]{wang16a}
Wang, H., Lu, Q., Huang, C., \& Wang, S. 2016, \apj, 821, 84,
  \dodoi{10.3847/0004-637X/821/2/84}

\bibitem[{{Warmuth} {et~al.}(2009){Warmuth}, {Holman}, {Dennis}, {Mann},
  {Aurass}, \& {Milligan}}]{2009ApJ...699..917W}
{Warmuth}, A., {Holman}, G.~D., {Dennis}, B.~R., {et~al.} 2009, \apj, 699, 917,
  \dodoi{10.1088/0004-637X/699/1/917}

\bibitem[{{Warren} \& {Warshall}(2001)}]{Warren2001}
{Warren}, H.~P., \& {Warshall}, A.~D. 2001, \apjl, 560, L87,
  \dodoi{10.1086/324060}

\bibitem[{{Yang} {et~al.}(2011){Yang}, {Cheng}, {Krucker}, \&
  {Hsieh}}]{Yang2011}
{Yang}, Y.-H., {Cheng}, C.~Z., {Krucker}, S., \& {Hsieh}, M.-S. 2011, \apj,
  732, 15, \dodoi{10.1088/0004-637X/732/1/15}

\bibitem[{{Yang} {et~al.}(2009){Yang}, {Cheng}, {Krucker}, {Lin}, \&
  {Ip}}]{Yang2009}
{Yang}, Y.-H., {Cheng}, C.~Z., {Krucker}, S., {Lin}, R.~P., \& {Ip}, W.~H.
  2009, \apj, 693, 132, \dodoi{10.1088/0004-637X/693/1/132}

\bibitem[{{Zharkova} \& {Gordovskyy}(2006)}]{2006ApJ...651..553Z}
{Zharkova}, V.~V., \& {Gordovskyy}, M. 2006, \apj, 651, 553,
  \dodoi{10.1086/506423}

\bibitem[{{Zhu} {et~al.}(2018){Zhu}, {Qiu}, \& {Longcope}}]{Zhu2018}
{Zhu}, C., {Qiu}, J., \& {Longcope}, D.~W. 2018, \apj, 856, 27,
  \dodoi{10.3847/1538-4357/aaad10}

\end{thebibliography}
\bibliographystyle{aasjournal}



\end{document}